\documentclass[journal]{IEEEtran}

\usepackage[T1]{fontenc}
\usepackage{upquote}

\usepackage{amssymb}
\usepackage{graphicx}
\usepackage{multirow}
\usepackage{color}
\usepackage{enumitem}
\usepackage{caption}
\usepackage{subcaption}
\usepackage{url}
\usepackage[cmex10]{amsmath}
\usepackage{booktabs}
\usepackage{array}

\usepackage{rotating}

\newcommand{\comment}[1]{}

\usepackage[ruled,noresetcount]{algorithm2e}
\SetKwBlock{KwMain}{main}{}
\SetKwBlock{KwInit}{initialize}{}
\SetKwBlock{KwFunction}{function}{}
\SetKwBlock{Begin}{begin}{end} 
\SetKwBlock{KwOn}{on}{}
\SetKwFunction{KwFuncSort}{Sort}
\SetKwFunction{KwFuncAvg}{DiffusDistConsensus}

\newcommand{\TRUE}{\ensuremath{ \textsc{true} }}
\newcommand{\FALSE}{\ensuremath{ \textsc{false} }}
\newcommand{\funcRecv}{\ensuremath{ \texttt{receive} }} 
\newcommand{\INITIATE}{\ensuremath{ \textsc{initiate} }}

\newcommand{\mb}[1]{\ensuremath{b_{#1}}}
\newcommand{\mA}[1]{\ensuremath{A_{#1}}}
\newcommand{\mz}[2]{\ensuremath{z_{#1}^{(#2)}}}
\newcommand{\msum}[2]{\ensuremath{\textit{sum}_{#1}^{(#2)}}}
\newcommand{\mave}[2]{\ensuremath{\hat{v}^{(#2)}}}

\newcommand{\mwa}[1]{\ensuremath{\overline{v}^{(#1)}}}
\newcommand{\mwe}[1]{\ensuremath{\hat{v}^{(#1)}}}
\newcommand{\me}[1]{\ensuremath{e_{#1}}}
\newcommand{\xh}[1]{\ensuremath{x^{(#1)}}}
\newcommand{\mx}[2]{\ensuremath{x_{#1}^{(#2)}}}
\newcommand{\mg}[2]{\ensuremath{\gamma_{#1}^{(#2)}}}

\newcommand{\Mp}[1]{\ensuremath{M_{#1}}}

\newcommand{\thresh}[1]{\ensuremath{{\mathcal{T}}_{_K}\left(#1\right)}}
\newcommand{\tp}{\ensuremath{^{\mathsf{T}}}}
\newcommand{\xopt}{\ensuremath{x^*}}
\newcommand{\cN}{\ensuremath{\mathcal{N}}}

\newcommand{\cf}{\ensuremath{f}}
\newcommand{\Rm}{\mathbb{R}}
\newcommand{\mR}{\mathbb{R}}
\newcommand{\fBound}{\ensuremath{B}}
\newcommand{\gBound}{\ensuremath{G}}
\newcommand{\hBound}{\ensuremath{H}}
\newcommand{\Lf}{\ensuremath{L_f}}
\newcommand{\Lfp}{\ensuremath{L_{f_p}}}
\newcommand{\eps}[1]{\ensuremath{\epsilon^{(#1)}}}

\newcommand{\Mk}[1]{\ensuremath{\mathcal{M}_{_K}\left(#1\right)}}

\newcommand{\Einf}{\ensuremath{E^{(\infty)}}}
\newcommand{\Et}[1]{\ensuremath{E^{(#1)}}}
\newcommand{\cBound}{\ensuremath{C}}
\newcommand{\ocBound}{\ensuremath{\overline{D}}}
\newcommand{\xs}{\ensuremath{x^*}}
\newcommand{\ms}[1]{\ensuremath{s^{(#1)}}}

\newcommand{\mv}[2]{\ensuremath{v_{#1}^{(#2)}}}
\newcommand{\wt}[3]{\ensuremath{w_{#1 #2}^{(#3)}}}
\newcommand{\vav}{\ensuremath{v_{av}}}
\newcommand{\Wt}[1]{\ensuremath{W^{(#1)}}}

\newcommand{\Ebt}[1]{\ensuremath{\overline{E}^{(#1)}}}

\newcommand{\DADMM}{\mbox{D-ADMM}}
\newcommand{\DIHT}{\mbox{DIHT}}
\newcommand{\CBDIHT}{\mbox{CB-DIHT}}

 \newcommand{\Lb}{L_{TV}}
 
 \newcommand{\Lmax}{L_{\textit{max}}}

\newcommand{\res}[2]{$#1${\tiny~$\times$~}$10^{#2}$}

\DeclareMathOperator*{\argmin}{arg\,min}

\newcommand{\headit}[1]{\emph{#1:}}

\newtheorem{theorem}{Theorem}[section]
\newtheorem{lemma}[theorem]{Lemma}
\newtheorem{proposition}[theorem]{Proposition}

\newtheorem{assumption}{Assumption}
\newtheorem{definition}{Definition}

\newenvironment{restate}[1]{\begin{trivlist} \item {\bf #1 (restated)} \em} {\end{trivlist}}

\begin{document}
%
\title{Distributed Compressed Sensing For Static and Time-Varying Networks}
%
%
%

\author{Stacy~Patterson~\IEEEmembership{Member,~IEEE,}
        	Yonina~C.~Eldar~\IEEEmembership{Fellow,~IEEE,}
	and~Idit~Keidar
\thanks{S. Patterson is with the Department of Computer Science, Rensselaer Polytechnic Institute, Troy, NY 12180 USA (email: sep@cs.rpi.edu).} 
\thanks{Y. C. Eldar, and I. Keidar are with the Department of Electrical Engineering, Technion~-~Israel Institute of Technology,~~Haifa 32000 Israel (email \{{yonina}, idish\}@ee.technion.ac.il).}
\thanks{The work of Y. C. Eldar is funded in part by the Intel Collaborative Research Institute for Computational Intelligence (ICRI-CI).  The work of I. Keidar is funded in part by the ICRI-CI, the Technion Autonomous Systems Program, and the Israeli Science Foundation.}}

%

\maketitle

\begin{abstract}
We consider the problem of in-network compressed sensing from distributed measurements.  Every agent has a set of measurements of a signal $x$, and the objective is for the agents to recover $x$ from their collective measurements using only communication with neighbors in the network.  Our distributed approach to this problem is based on the centralized Iterative Hard Thresholding algorithm (IHT).  We first present a distributed IHT algorithm for static networks that leverages standard tools from distributed computing to execute in-network computations with minimized bandwidth consumption.  Next, we address distributed signal recovery in networks with time-varying topologies.  The network dynamics necessarily introduce inaccuracies to our in-network computations.  To accommodate these inaccuracies, we show how centralized IHT can be extended to include inexact computations while still providing the same recovery guarantees as the original IHT algorithm.  We then leverage these new theoretical results to develop a distributed version of IHT for time-varying networks.  Evaluations show that our distributed algorithms for both static and time-varying networks outperform previously proposed solutions in time and bandwidth by several orders of magnitude.
\end{abstract}


%
\IEEEpeerreviewmaketitle

\section{Introduction} \label{intro.sec}

\IEEEPARstart{I}{n} compressed sensing, a sparse signal $x \in \mR^N$ is sampled and compressed into a set of $M$ measurements, where $M$
is typically much smaller than $N$.  If these measurements are taken appropriately, then it is possible to recover $x$
from  this small set of  measurements using a variety of polynomial-time algorithms~\cite{EK12}. 

Compressed sensing is an appealing approach for sensor networks, where measurement capabilities may be limited 
due to both coverage and energy constraints. Recent works have demonstrated that compressed sensing is applicable to a variety of 
sensor networks problems including event detection \cite{MLH09}, urban environment monitoring \cite{LZZL11} and traffic estimation \cite{YZZ10}. 
In these applications, measurements of the signal are taken by sensors that are distributed throughout a region.
The measurements are then collected at a single fusion center where signal recovery is performed.  
While the vast majority of recovery algorithms consider a centralized setting,  a centralized approach is not always feasible, especially in sensor networks where no powerful computing center is available and where bandwidth is limited. 

Since the measurements are already distributed throughout the network, it is desirable to perform the signal recovery within the network itself.
Distributed solutions for compressed sensing have begun to receive attention lately~\cite{BG10,MXAP11,MXAP12,MXAP13}.  
Although these algorithms converge to a correct solution, they do not optimize for metrics that are important in a distributed setting, most notably, bandwidth consumption.   
In addition, these techniques often have a high computational cost as they require every agent to solve a convex optimization problem in each iteration.  Such computational capacity may not be available in low-power sensor networks.

We propose an  alternative approach to distributed compressed sensing that is based on \emph{Iterative Hard Thresholding} (IHT)~\cite{BD09}.
In a centralized setting, IHT offers the benefit of computational simplicity when compared to methods like basis pursuit. 
Our distributed approach maintains this same computational benefit.  In addition, recent work~\cite{BE12} has established that centralized IHT
can be used for problems beyond compressed sensing, for example sparse signal recovery from nonlinear measurements.  Our distributed solution
provides the same recovery guarantees as centralized IHT and thus can also be applied to these settings.

In our distributed implementation of IHT, which we call \DIHT, all agents store identical copies of an estimate  of $x$.  
In each iteration, every agent first performs a \emph{simple local computation} 
to derive an intermediate vector.  The agents then perform a \emph{global computation} on their intermediate vectors to derive the next iterate.
This global computation is performed using only communication between neighbors in the network.

We present two versions of our distributed algorithm, one for static networks and one for networks with time-varying topologies. 
In the version for static networks, we employ standard tools from  distributed computing to perform the global computation in a simple, efficient manner.
The result is a distributed algorithm that outperforms previous solutions in both bandwidth and time by several orders of magnitude.

In networks that are time-varying, it is not possible to perform the global computation exactly unless each agent has a priori knowledge of the network dynamics.  However, it is possible to approximate the global computation using only local communication.  
We first  show how  centralized IHT can be extended to accommodate inexact computations while providing the same recovery guarantees as
the original IHT formulation.  We then leverage these new theoretical results to develop a version of \DIHT\ that uses multiple rounds of a distributed consensus algorithm~\cite{T84} to execute each inexact global computation.  We call this algorithm \emph{consensus-based \DIHT}, or \CBDIHT.  Evaluations show that CB-DIHT requires several orders of magnitude less time and bandwidth than the best-known, previously proposed solution.

\subsection{Related Work} \label{related.sec}

Several recent works have proposed distributed algorithms that can apply to a basis pursuit formulation 
of the distributed compressed sensing problem.
In these, the signal is recovered by solving a convex formulation of the original recovery problem. 
These distributed methods can be divided into two classes:  double-looped algorithms and single-looped algorithms.

The double looped techniques~\cite{MXAP09,MXAP11,JXM11} consist of an inner loop, in which agents solve a dual problem, and an outer loop where the Lagrange multipliers are updated locally.  In each iteration of the inner loop, the agents exchange $N$-vectors with their neighbors, and multiple inner loop iterations are needed to solve the dual problem.  With the exception of~\cite{JXM11}, these algorithms
all require a static network.
In single-looped methods~\cite{BG10,MXAP12,MXAP13}, in each iteration, every agent solves a local convex optimization problem;  it also exchanges an $N$-vector 
with each of its neighbors and uses this vector to update the parameters of its local optimization problem.
A recent work~\cite{MXAP12} presented an experimental evaluation of these methods and demonstrated that  the single-looped algorithm 
\DADMM~\cite{MXAP12,MXAP13} outperformed the other algorithms. 
While \DADMM\ uses only local communication, each agent must send its  entire estimate vector to every neighbor in every iteration.   This vector may not be sparse for many iterations, and therefore, bandwidth usage can be high.  
Furthermore, the convergence time increases as the network connectivity increases, whereas the convergence rates of \DIHT\ and \CBDIHT\  improve with increased network connectivity.
We note that the convergence of \DADMM\ has only been established theoretically for bipartite graphs, but experiments have demonstrated convergence in general graphs.

The distributed subgradient algorithm~\cite{NO09,LOF11} was proposed as a general distributed convex optimization technique but can be adapted to basis pursuit.
In this approach, every agent stores an estimate of the signal.  In each iteration, it exchanges its estimate with its neighbors and then performs a local projected subgradient step.  
The algorithm converges in static and time-varying networks, though the convergence rate can be slow. Simulations have shown that, in a time-varying graph, the distributed subgradient method converges more quickly than the double looped algorithm in~\cite{JXM11}.

The work by Ravazzi et al.~\cite{RFM13} proposes a distributed algorithm based on iterative soft thresholding.  This algorithm is similar to \DIHT\ for static networks, however, it only converges in complete, static graphs.  This is in contrast with \DIHT\ which converges in any connected, static graph.

Our approach for \CBDIHT\ was inspired by recent work on a distributed proximal gradient algorithm~\cite{CO12}.
This algorithm also simulates a centralized, inexact optimization method, in this case, a proximal gradient method, 
and it uses multiple iterations of distributed consensus to perform each inexact computation.
This work differs from ours in  that the convergence of the inexact centralized proximal gradient method had already been established~\cite{SRB11},
whereas no such prior analysis exists for centralized IHT.  
Furthermore, the distributed proximal gradient algorithm depends on assumptions that
are not compatible with standard compressed sensing formulations such as the one in this paper.

The convergence of centralized IHT was established in~\cite{BD08}, and its application to compressed sensing  was studied in~\cite{BD09}.
Recently, Beck and Eldar adapted IHT to signal recovery for more general nonlinear objective functions
and provided theoretical guarantees on signal recovery in this setting~\cite{BE12}. We leverage this work in our distributed algorithms.
A variation on IHT for nonlinear measurements was also proposed in~\cite{B13}.  This work uses a Taylor series approximation for the gradient step in each iteration rather than an exact gradient as in~\cite{BE12}.

Finally, we note that, in a related work~\cite{PEK13}, we present an extension to \DIHT\ for static networks that can further reduce bandwidth for problems that require many rounds to converge.

\subsection{Outline}
The remainder of the paper is organized as follows.  In Section~\ref{problem.sec}, we detail our problem setting and formulation.
In Section \ref{staticalg.sec}, we present the \DIHT\ algorithm for static networks, and in Section \ref{tvalg.sec}, we present the \CBDIHT\ algorithm for time-varying networks. 
Section~\ref{eval.sec} provides numerical results demonstrating the performance of \DIHT\ and \CBDIHT. A 
discussion on the fault tolerance and recovery guarantees of distributed compressed sensing algorithms is given in Section~\ref{discussion.sec}. 
Concluding remarks are provided in Section \ref{conclusion.sec}.


%


\section{Problem Formulation}
\label{problem.sec}

We consider a network of $P$ agents.   The agents may be sensors themselves or they may be fusion nodes that collect measurements from several nearby sensors. We assume there is a unique agent identified as agent 1. 
This agent can be chosen using a variety of well-known distributed algorithms (see~\cite{L96}).

The agents seek to estimate a signal $x \in \mR^N$ that is $K$-sparse, meaning $x$ has at most $K$ non-zero elements. 
Each agent has one or more (possibly noisy) measurements of the signal, and each has a loss function ${\cf_p:\Rm^N\rightarrow\Rm}$, known only to agent $p$,
that indicates how well a given vector satisfies its measurements.  
The goal is for every agent to recover $x$ from their collective measurements using only communication between neighbors in the network.
To recover $x$, the agents attempt to solve the following optimization problem,
\begin{equation}
\text{minimize}~\cf(x) := \sum_{p=1}^{P} \cf_p(x)~~\text{subject to}~~\|x\|_0 \leq K,
 \label{opt.eq}
\end{equation}
where $\| \cdot \|_0$ denotes the $\ell_0$ psuedo-norm, i.e., the number of non-zero components.
Note that each agent only has access to its own measurements, and so the agents must collaborate to solve the optimization problem.

The following assumption is made throughout the paper.
\setcounter{assumption}{-1}
\begin{assumption} \label{k.assum}
Agent 1 knows the sparsity parameter $K$ of the signal $x$ to be estimated.
\end{assumption}
We note that this assumption is made only for convenience. In practice, IHT can be implemented without this knowledge by only keeping the elements above a threshold.

We also make the following assumptions about the loss functions.
\begin{assumption}
\label{f.assum}
The loss functions $\cf_p$, $p=1 \ldots P$, satisfy the following conditions:
\begin{enumerate}[label=(\alph*)]

\item There exists a $\fBound_p \in \mR$ such that for all $x \in \mR^N, \cf_p(x) \geq \fBound_p$. 
\item The gradient $\nabla \cf_p$  is Lipschitz continuous over $\mR^N$ with Lipschitz constant $\Lfp$, i.e.,
\[
\| \nabla \cf_p(x) - \nabla \cf_p(y) \|_2 \leq \Lfp \| x - y \|_2, ~~~~ \forall x,y \in \mR^N.
\]
Note that this implies that $\cf(x) = \sum_{p=1}^P \cf_p(x)$ is also Lipschitz continuous over $\mR^N$.
\item There exists $\gBound_p, \hBound_p \in \mR$ such that for all $x \in \mR^N$,
\[
\| \nabla \cf_p(x)\|_2 \leq \gBound_p \| x \|_2 + \hBound_p.
\] 
\item Every agent $p$ knows its $\Lfp$, and agent 1 knows an upper bound on 
the Lipschitz constant $\Lf$ for $\cf(x) = \sum_{p=1}^P \cf_p(x)$.  
\end{enumerate}
\end{assumption}
The agents do not know $\fBound_p$, $\gBound_p$, nor $\hBound_p$. 

If agent 1 does not know an upper bound on $\Lf$, then
it can learn one  using a distributed algorithm.  One such upper bound is $\sum_{p=1}^P \Lfp$.  
Distributed algorithms for computing an upper bound for $\Lf$ are described in Appendix~\ref{distL.app}.

As a specific example problem, we consider compressed sensing~\cite{EK12}.  
Here, each agent $p$ has $\Mp{p}$ linear measurements of $x$ taken using its sensing matrix $\mA{p} \in \mR^{\Mp{p} \cdot N}$.
The measurement vector of agent $p$, denoted $\mb{p}$, is given by $\mb{p} = \mA{p} x + \me{p}$,
where $\me{p} \in \mR^{\Mp{p}}$ is the measurement error for agent $p$.
The loss function for each agent  is  $\cf_p(x) := \| \mA{p} x - \mb{p} \|_2^2$.
It is straightforward to verify that this loss function satisfies Assumption \ref{f.assum}. 

We define 
\[
f(x) = \sum_{p=1}^P \| \mA{p} x - \mb{p} \|_2^2 = \| A x - b\|_2^2,
\]
where
\[
\renewcommand*{\arraystretch}{1.3}
A :=  \left[ \begin{array}{c}
~~~~~~~~\mA{1}~~~~~~~~~ \\
\hline 
\vdots \\
\hline
\mA{P}
\end{array} \right], ~~~~~~~~ b := \left[ \begin{array}{c}
\mb{1} \\
\hline
\vdots \\
\hline
\mb{P} 
\end{array} \right].
\] 
The objective is for agents to collaborate to solve the compressed sensing problem,
\begin{equation}
\text{minimize}~\| A x - b\|_2^2~~~\text{subject to}~~\|x\|_0 \leq K. 
 \label{optcs.eq}
\end{equation}

In the sequel, we present solutions for the proposed \emph{distributed sparse signal recovery problem} for two different network models, a static network and a time-varying network.

\headit{Static Network Model}  We model the network by an undirected, connected graph.  Agents can communicate only with their neighbors in the graph.  Messaging is reliable but asynchronous, meaning that every message that is sent is eventually delivered, but the delay between sending and delivery may be arbitrarily long.

\headit{Time-Varying Network Model}
Here, we consider a discrete time model. 
 At each time  step $t$, the network is modeled by a directed graph $(V, \Et{t})$, where $V$ is the set of $P$ agents and $\Et{t}$ are the directed communication links between them at time $t$.  If $(q,p) \in \Et{t}$, then agent $p$ can send a message to agent $q$ in time step $t$.  Messaging is reliable and synchronous, meaning that any message sent in time $t$ is received before time $t+1$.
 We adopt the following standard assumption about the network connectivity over time~\cite{BHOT05,NO09,CO12}.  
  \begin{assumption} \label{network.assum}
The sequence of graphs $(V, \Et{t})$, $t= 0,1,2, \ldots$, satisfies the following conditions: 
\begin{enumerate}[label=(\alph*)]
\item The graph $(V, \Einf)$ is strongly connected, where $\Einf$ is the set of edges that appear in infinitely many time steps.  
\item There exists an integer $\cBound \geq 1$ such that if $(q,p) \in \Einf$, then
$(q,p) \in \Et{t} \cup \Et{t+1} \cup \cdots \cup \Et{t+ \cBound-1}$, for all~$t~\geq~0$.

\end{enumerate}
\end{assumption}
In short, this assumption means that, while the network may not be connected at any given time step, the union of graphs over each interval of $\overline{\cBound} = (P-1)\cBound$ time steps is a strongly connected graph.  The agents do not know the value of $\cBound$. 

Our goal, in both network settings, is for the agents to recover the same sparse signal from their private loss functions using only local communication.
In Section \ref{staticalg.sec}, we present our distributed recovery algorithm for static networks.  In Section \ref{tvalg.sec}, we extend this algorithm to the time-varying networks. 


\vspace{-.2cm}
\section{Distributed Algorithm for Static Networks}
\label{staticalg.sec}
Problem (\ref{opt.eq}) is known to be NP-Hard in general \cite{N95}.  However, for suitable loss functions, efficient centralized algorithms exist.
Our distributed recovery algorithm is based on 
Iterative Hard Thresholding~\cite{BD08,BD09}.
We first briefly review this method and related convergence results.  We then provide the details of our distributed algorithm.
 
\vspace{-.2cm}
\subsection{Iterative Hard Thresholding}
\label{sec:algorithm}
Consider a $K$-sparse signal $\xs$ that has been measured and a loss function $\cf: \mR^N \rightarrow \mR$ that captures how well a given vector matches those measurements.
We assume that $\cf$ is lower bounded and that it has a  Lipschitz-contiuous gradient with constant $\Lf$.
IHT~\cite{BD08,BD09,BE12}  is a gradient-like, centralized algorithm that recovers $\xs$ by solving the optimization problem,
\begin{equation} \label{centralopt.eq}
\text{minimize}~\cf(x) ~~\text{subject to}~~\|x\|_0 \leq K.
\end{equation}

Let  $\thresh{v}$ be the thresholding operator which returns a vector where all but the $K$ entries of $v$ with the largest magnitude are set to 0 (with ties broken arbitrarily).	
IHT begins with an arbitrary $K$-sparse vector $\xh{0}$.
In each iteration, a gradient-step is performed, followed by application of the thresholding operator.  This iteration is given by,
\begin{equation}
\xh{k+1}= \thresh{\xh{k} - \textstyle \frac{1}{L} \nabla \cf(\xh{k})},
\label{iht.eq}
\end{equation}
where $L > \Lf$ is a constant.

The loss function $\cf$ is not necessarily convex, and, in general, optimization algorithms for non-convex objective functions only guarantee convergence to a stationary point.  The inclusion of the sparsity constraint, which is also non-convex, means that we cannot employ the same definition of stationarity that is used for problems with convex constraints.  We instead use a definition of a stationary point that is relevant to problem (\ref{centralopt.eq}) called \emph{$L$-stationarity} (see \cite{BE12} for details). 
\begin{definition} \label{lstation.def}
For a given $L > 0$, a $K$-sparse vector $\xs \in \mR^N$ is an \emph{$L$-stationary point} of problem (\ref{centralopt.eq}) if it satisfies,
\[
\xs = \thresh{\xs - \textstyle \frac{1}{L} \nabla \cf(\xs)}.
\]
\end{definition}
It has been shown that $L$-stationarity is a necessary condition for optimality (see Thm. 2.2 in~\cite{BE12}).

With this definition, we can now state the relevant convergence result for IHT with a general nonlinear objective.
\begin{theorem}[Thm. 3.1 in~\cite{BE12}]
Let $\cf$ be lower-bounded and let $\nabla \cf$ be Lipschitz-continuous with constant \Lf.  Let $\{\xh{k}\}_{k \geq 0}$ be the sequence generated by IHT with  $L >  \Lf$.
Then, any accumulation point of   $\{\xh{k}\}_{k \geq 0}$ is an $L$-stationary point of (\ref{centralopt.eq}) .
\end{theorem}
We note that this theorem does not guarantee that IHT will converge to an $L$-stationary point; it only guarantees that if the algorithm converges,
the accumulation point is an $L$-stationary point.  More details on the convergence behavior of IHT for nonlinear objectives can be found in~\cite{BE12}.

For the compressed sensing problem, a stronger result has been shown.
\begin{theorem}[Thm. 5 in~\cite{BD09}]
Let $\xs$ be a $K$-sparse signal sampled with error $e$, i.e., $b = A \xs + e$.  Let $\| A \|_2  < 1$, and let $A$ satisfy the restricted isometry property~\cite{C06}  with $\delta_{3K} < \frac{1}{\sqrt{32}}$. Then the sequence $\{\xh{k}\}_{k \geq 0}$ generated by IHT with $L=1$ satisfies
\[
\|\xh{k} - \xs \|_2 \leq 2^{-k}\|\xs \|_2 + 5 \| e \|_2.
\]
\end{theorem}
This theorem implies that, if the measurements are taken without error, then IHT recovers the original signal.

\subsection{Distributed Iterative Hard Thresholding} \label{dihtdesc.sec}

We now present our distributed implementation of IHT for static networks.
Every agent stores an identical copy of the signal estimate $\xh{k}$.
In iteration $k$, each agent first performs a local computation
to derive an intermediate vector $\mz{p}{k}$.  The agents then perform a global computation on their intermediate vectors to derive the next iterate
$\xh{k+1}$, which is, again, identical at every agent.
We now define these computations.

\vspace{.1cm}
\noindent \textit{Local computation:}  Agent $p$ computes its intermediate vector,
\begin{equation}
\mz{p}{k} =  \nabla \cf_p(\xh{k}), \label{eq:z}
\end{equation}
using its local loss function and the current iterate $\xh{k}$.
This vector can be computed using only local information.

\vspace{.1cm}
\noindent \textit{Global computation:}
In the global computation step, all agents must compute a function $G$ that depends on all of their intermediate vectors. This function is defined as follows,
\begin{equation}
\xh{k+1} = G(\mz{1}{k}, \ldots, \mz{P}{k}) :=  \thresh{\xh{k} - \textstyle \frac{1}{L}  \sum_{p=1}^P \mz{p}{k}}. \label{global.eq} 
\end{equation}
To find $G$, first, the agents compute the sum of their intermediate vectors
using a well-known distributed algorithm called \emph{broadcast/convergecast}~\cite{S83} (described below). 
This sum is then used to complete the gradient step, followed by application of the threshold operator.  
The combination of the local computation (\ref{eq:z}) and the global computation (\ref{global.eq}) is equivalent to one iteration of centralized IHT 
in (\ref{iht.eq}). 

We now describe DIHT.
The agents first create a breadth-first spanning tree over the network, rooted at agent 1, using a distributed algorithm (see~\cite{AW04} for details). 
This requires $2|E| - (P-1)$ messages, where $|E|$ is the number of edges in the network. 
After the tree is constructed, each agent knows the IDs of its parent and its children in the spanning tree. 
The creation of this tree can be done as a pre-processing step, but in practice, the tree creation is usually done in conjunction with the first broadcast.

Agent 1 initializes its estimate vector $\mx{1}{0}$ to $x_{\textit{init}}$.  
In each iteration $k$,  agent 1 computes its intermediate vector $\mz{1}{k}$ according to (\ref{eq:z}).  It then sends $\mx{1}{k}$ to its children.  On receipt of $\mx{1}{k}$ from its parent, an agent updates its own estimate $\mx{p}{k}$ to equal $\mx{1}{k}$.  It then computes $\mz{p}{k}$ by (\ref{eq:z}).   The agent sends $\mx{1}{k}$ to its children, if it has any.  If an agent does not have any children, it sends its vector $\mz{p}{k}$ to its parent.
Once an agent has received vectors from all of its children, it adds those vectors to its $\mz{p}{k}$ and sends the resulting sum to its parent (if it is not agent 1).  
When agent 1 receives vectors from all of its children, it adds these vectors to $\mz{1}{k}$ and finishes the global computation (\ref{global.eq}) to obtain $\mx{1}{k+1}$.  This completes one iteration.  The process is then repeated to obtain the next iterate.
Pseudocode for \DIHT\ is given in Algorithm \ref{static_diht1.alg}.

\begin{algorithm}[t]  
\caption{DIHT}  \small
\label{static_diht1.alg}
\SetAlgoNoLine
\SetAlgoNoEnd
\DontPrintSemicolon
\SetNoFillComment
\KwInit{
$k \gets 0$ \;
$\mx{1}{0} \gets x_{\textit{init}}$ \;
}
\BlankLine
\emph{Algorithm executed by agent 1.} \;
\While{\TRUE}{
$\mz{1}{k} \gets \nabla \cf_1(\mx{1}{k})$ \;
Send $\mx{1}{k}$ to children \;
Receive $\msum{q}{k}$ from each $q \in \mbox{children}(1)$ \;
$\msum{1}{k} \gets \left(\sum_{q \in \mbox{children}(1)} \msum{q}{k}\right)~+~\mz{1}{k}$  \;
$\mx{1}{k+1} \gets \thresh{\mx{1}{k} - \textstyle \frac{1}{L} \msum{1}{k}}$ \;
$k \gets k+1$ \;
}
\BlankLine 
\BlankLine
\emph{Algorithm executed by agent $p \neq 1$.} \;
\KwOn({$\funcRecv_p(\mx{1}{k})$} from parent){ 
	$\mx{p}{k} \gets \mx{1}{k}$ \;
	$\mz{p}{k} \gets \nabla \cf_p(\mx{p}{k})$\;
	Send $\mx{1}{k}$ to children \;
	Receive $\msum{q}{k}$ from each $q \in \mbox{children}(p)$ \;
	$\msum{p}{k} \gets \left(\sum_{q \in \mbox{children}(p)} \msum{q}{k}\right)~+~\mz{p}{k}$   \;
	Send $\msum{p}{k}$ to parent \;
	$k \gets k+1$ \;
}
\end{algorithm}

\subsection{Algorithm Analysis}
\DIHT\ requires $O(N)$ storage at each agent.  In every iteration, every agent computes the gradient of its local loss function. For compressed sensing, this requires only matrix-vector multiplication.  Therefore, the local computation is much simpler than the double and single looped algorithms that require each agent to solve a convex optimization problem in every iteration.  An iteration of \DIHT\ consists of a broadcast in which a $K$-sparse vector is sent down the tree and a convergecast in which the intermediate $N$-vectors are aggregated up the tree.  As the tree has $P-1$ edges,
$2(P-1)$ total messages are sent per iteration.   For the broadcast, each agent sends at most $D_p - 1$ messages, where $D_p$ is the node degree in the original graph.  For the convergecast, the agent sends a single message to its parent in the tree, for a total of $D_p$ messages per agent per iteration.
While agent 1 plays a unique role in the global computation, it performs the same 
number and types of computations as every other agent, with the exception of the thresholding operation which requires  a single scan of the sum vector.  One approach to find the $K$ largest magnitude values in a single scan is for agent 1 to keep a priority queue, initially containing the first $K$ components of the sum vector.
 Starting with component $K+1$, the agent checks each remaining component in the sum vector against the smallest entry in the queue.  If the component is larger, then the component is added to the queue and the smallest entry is removed.  Each enqueue or dequeue operation has $\log(K)$ time complexity, and  checking the smallest entry can be done in constant time.  
Thus, the running time of this approach is $O(N\log(K))$.

The estimates $\mx{p}{k}$, $p=1 \ldots P$, are equivalent to each other in all iterations, and they evolve exactly as $\xh{k}$ in (\ref{global.eq}). 
Thus, \DIHT\ provides the same convergence guarantees as centralized IHT.  This is formalized in the following theorems.
\begin{theorem}
Let each $\cf_p$, $p=1 \ldots P$, satisfy Assumption \ref{f.assum}, and let $\{\mx{p}{k}\}_{k \geq 0}$, $p=1 \ldots P$, be the sequences of estimates generated by \DIHT\ with  $L >  \Lf$  in a static network.
Then, any accumulation point of $\{\mx{p}{k}\}_{k \geq 0}$, $p=1 \ldots P$, is an $L$-stationary point of (\ref{opt.eq}) .
\end{theorem}

\begin{theorem} \label{dihtconverge.thm}
For the distributed compressed sensing problem (\ref{optcs.eq}), let $\xs$ be the original $K$-sparse signal measured with error $e$, and let $A$ be such that  $\| A \|_2  < 1$ and satisfy the restricted isometry property  with $\delta_{3K} < \frac{1}{\sqrt{32}}$. Then, the sequences of estimates $\{\mx{p}{k}\}_{k \geq 0}$, $p=1 \ldots P$, generated by \DIHT\ with $L=1$ in a static network  satisfy 
\[
\|\mx{p}{k} - \xs \|_2 \leq 2^{-k}\|\xs \|_2 + 5 \| e \|_2.
\]
\end{theorem}


\section{Distributed Algorithm for Time-Varying Networks} \label{tvalg.sec}

We now show how to extend  DIHT to networks with time-varying topologies. 
The local computation step (\ref{eq:z}) remains the same.  For the global computation step (\ref{global.eq}), the broadcast-convergecast algorithm used  to compute a sum in \DIHT\ requires a static network; it cannot be applied in a time-varying network setting.  In fact, without a priori knowledge of the network dynamics or membership, it is not possible for the agents to perform the global computation in finite time using \emph{any} algorithm. This is because, without this knowledge, an agent cannot determine when it has received the information it needs (from all other agents)  to compute the sum in (\ref{global.eq}).

In our extended DIHT algorithm, we use a distributed consensus algorithm~\cite{T84} to \emph{approximate} the average of the intermediate vectors and then use this approximation to complete the gradient step, followed by application of the thresholding operator.  
Distributed consensus can be implemented without any global knowledge of the network, and  it has been shown that, in time-varying networks,  distributed consensus converges to the average of the agents' initial values~\cite{BHOT05}.  After a finite number of 
iterations, the agents learn an approximation of this average.  To use distributed consensus for the global computation steps in DIHT, we must first consider the effects of such approximation errors on centralized IHT.  

In Section \ref{ihtin.sec}, we present new theoretical results on the convergence of 
 centralized IHT  with approximate sums.  We capture these approximations in the form of inexact computations of $\nabla \cf$.  We show that, under a limited assumption on the accuracy of the gradient values, IHT with inexact gradients provides the same recovery guarantees as IHT with exact gradient computations. 
 We then  leverage these new theoretical results to develop a consensus-based distributed IHT algorithm for time-varying networks.  

\subsection{IHT with Inexact Gradients} \label{ihtin.sec}
IHT with inexact gradients is identical to IHT in (\ref{iht.eq}), except that, in each iteration the gradient is computed approximately.
The iteration is thus given by,
\begin{equation}
\xh{k+1} = \thresh{\xh{k} - \textstyle \frac{1}{L} \left(\nabla f(\xh{k}) + \eps{k} \right)}.
\label{ihtin.eq}
\end{equation}
Here, $\eps{k} \in \mR^N$ is the error in the gradient computation in iteration $k$. 

The following theorems show that, so long as the sequence $\{ \|\eps{k} \|^2\}_{t \geq 0}$ is summable,  algorithm (\ref{ihtin.eq}) provides the same convergence guarantees as IHT with exact gradients. Proofs are deferred to Appendix~\ref{proofs.app}.
Our first theorem (analogous to Theorem 3.1 in \cite{BE12}) states that any accumulation point is an $L$-stationary point (as defined in Definition \ref{lstation.def}).
\newcommand{\thmLPoint}{
Let $\cf$ be lower-bounded and let $\nabla \cf$ be Lipschitz-continuous with constant \Lf.  Let $\{\xh{k}\}_{k \geq 0}$ be the sequence generated by (\ref{ihtin.eq})   with  $L >  \Lf$ and with a sequence $\{\eps{k}\}_{k \geq 0}$ satisfying $\sum_{k=0}^{\infty} \| \eps{k} \|_2^2 < \infty$.  Then, any accumulation point of  $\{\xh{k}\}_{k \geq 0}$ is an $L$-stationary point.
}
\begin{theorem} \label{lstation.thm}
\thmLPoint
\end{theorem}

For the compressed sensing problem, we can show a stronger result (analogous to Theorem 3.2 in~\cite{BE12}).
Let $f(x) = \|A x - b \|_2^2$. 
The spark of $A$, denoted $\text{spark}(A)$ is the smallest number of columns of $A$ that are linearly dependent.
If $\text{spark}(A) > K$,  then algorithm (\ref{ihtin.eq}) converges to an $L$-stationary point.
\newcommand{\thmCS}{
Let $f(x) = \|A x - b \|_2^2$, with $\text{spark(A)} > K$. Let $\{\xh{k}\}_{k\geq 0}$ be the sequence generated by (\ref{ihtin.eq})  
with $L > 2\lambda_{max}(A\tp A)$ and with a sequence $\{\eps{k}\}_{t \geq 0}$ satisfying $\sum_{k=0}^{\infty} \| \eps{k} \|_2^2 < \infty$. 
Then, the sequence $\{\xh{k}\}_{k\geq 0}$ converges to an $L$-stationary point. 
}

\begin{theorem}\label{lconverge.thm}
\thmCS
\end{theorem}

\subsection{Distributed Diffusive Consensus} \label{distav.sec}
As previously stated, in each iteration of CB-DIHT, the agents use distributed consensus to compute an approximation of the average of their intermediate vectors $\mz{p}{k}$, $p=1 \ldots P$.  
In the standard formulation of  distributed consensus in a time-varying network,
every agent has an initial, vector-valued state $\mv{p}{0}$.  In time step $t$, every agent computes a weighted average of its value and that of its neighbors in that time step.  The vector at agent $p$ evolves as,
\begin{equation} \label{distav.eq}
\mv{p}{t+1} = \sum_{q=1}^P \wt{p}{q}{t}~\mv{q}{t},
\end{equation}                                                       
where $\wt{p}{q}{t}$ is the weight that agent $p$ assigns to the value at agent $q$.
 Under appropriate assumptions about the weights and the network connectivity over time (e.g.,~Assumption \ref{network.assum}), the agents' vectors converge to $\vav = \frac{1}{P} \sum_{p=1}^P \mv{p}{0}$~\cite{BHOT05}.

In \CBDIHT, the agents need to compute an approximate average in each iteration.  As with DIHT for static networks,  agent 1 initiates this global computation and computes the next iterate once the global computation is complete.  To use distributed consensus in the global computation step, we must augment the standard consensus algorithm so that it can be initiated by a single agent, just as agent 1 initiated the broadcast/convergcast algorithm in \DIHT\ for static networks.  We now explain the details of our modified consensus algorithm, which we call \emph{diffusive distributed consensus}.

The algorithm operates in discrete time steps.  In any step, an agent may be \emph{initiated}, meaning it is participating in the consensus algorithm, or it may be \emph{uninitiated}, meaning it is not yet participating in the algorithm. We call a link $(q,p)$ \emph{active} at time $t$ if agents $p$ and $q$ were initiated prior to time $t$.  
We assume that agent 1 begins the algorithm at time step 0 and thus is the only initiated agent at that time.  In step 0, agent 1 sends an \INITIATE\ message along its outgoing links in that time step, $i.e.$, it sends messages to all agents $p$ such that $(p,1) \in \Et{0}$.  Upon receipt of this message, an agent is initiated. In time step 1, the initiated agents begin the consensus algorithm specified by (\ref{distav.eq}) over the active links that are present in that time step.  
If agent $p$ is initiated in time step $t$, in all steps $T>t$, it sends \INITIATE\ messages over any adjacent, inactive links, thus activating them and initiating those adjacent nodes if necessary.   The agent also performs the consensus iteration (\ref{distav.eq}) over its active links in each time step $T>t$.  In this manner, the \INITIATE\ message diffuses through the network until the entire network is participating in the consensus algorithm, at which point the algorithm is identical to standard distributed consensus.  
Pseudocode for the diffusive distributed consensus algorithm is given in Appendix~\ref{diffcon.app}.
One can think of diffusive distributed consensus as a standard consensus algorithm over a graph $(V, \Ebt{t})$, 
where $\Ebt{t} \subseteq \Et{t}$ contains only active links.

In a time-varying network, an agent may not receive the \INITIATE\ message containing $x_1^{(k)}$ in every iteration.
Furthermore, it may receive an \INITIATE\ message for  $x_1^{(j)}$, with $j<k$, after it receives $x_1^{(k)}$.
The iteration number $k$ is included in each \INITIATE\ message so that an agent can determine
whether the message contains the most up-to-date iterate it has seen so far. If it does, then the agent
uses this iterate to compute its intermediate vector and begins the consensus algorithm with its active neighbors.  Otherwise, the agent ignores the message.

To ensure convergence of the diffusive distributed consensus algorithm, we require that the time-varying network satisfy the network connectivity conditions in Assumption \ref{network.assum}.  
A consequence of this assumption is that, after at most $C(P-1)$ time steps, every agent is activated.
Therefore,  for $\ocBound ={2(P-1)\cBound}$, we have $\Ebt{t} \cup \Ebt{t+1} \cup \cdots \cup \Ebt{t + \ocBound -1} = \Einf$.

We also make the following standard assumption on the weights used in iteration (\ref{distav.eq})~\cite{NO09,BHOT05}. 
\begin{assumption} \label{weight.assum}
The weight matrices $W(t) := [\wt{p}{q}{t}]$, $t = 0, 1, 2, \ldots$ satisfy the following conditions:
\begin{enumerate}[label=(\alph*)]
\item The matrix $\Wt{t}$ is doubly stochastic.
\item There exists a scalar $\eta \in (0,1)$, such that for all $p$,~ ~$\wt{p}{p}{t}\geq\eta$. 
Further, if $(p,q) \in \Ebt{t}$, then $\wt{p}{q}{t} \geq \eta $, and if $(p,q) \notin \Ebt{t}$, then $\wt{p}{q}{t} = 0$.
\end{enumerate}
\end{assumption}

Under Assumptions  \ref{network.assum} and \ref{weight.assum}, we can bound the deviation between the average $\vav$ and any agent's
estimate of that average after $s$ time steps of diffusive distributed consensus.  
\begin{proposition} \label{avg.thm}
Let the network satisfy Assumption \ref{network.assum}.  After $s$ time steps of diffusive distributed consensus, initiated at a single agent, where the weights obey Assumption \ref{weight.assum},
the deviation of each agent's estimate from the average $\vav$ satisfies,
\[
\| \mv{p}{s} - \vav \|_2 \leq \Gamma \gamma^s \sum_{q=1}^P \|\mv{q}{0} \|_2,~~  \text{for}~p=1, \ldots P,
\]
where $\Gamma = 2 (1 + \eta^{-\overline{D}})/(1 - \eta^{\overline{D}})$ and $\gamma = (1 - \eta^{\overline{D}})^{1/\overline{D}}$,
with $\ocBound = 2(P-1) \cBound$.
\end{proposition}
This proposition is a straightforward extension of Proposition~1 in~\cite{NO09} for the standard distributed consensus algorithm.  We therefore omit the proof for brevity.

\subsection{Consensus-Based DIHT} \label{tvdiht.sec}

\begin{algorithm}[t]
\caption{Consensus-Based DIHT.}  \small
\label{tvdiht.alg}
\SetAlgoNoLine
\SetAlgoNoEnd
\DontPrintSemicolon
\SetNoFillComment
\KwInit{
$\mx{p}{0} \gets x_{\textit{init}}$ \;
$k \gets 0$ \;
}
\BlankLine
\BlankLine
\emph{Algorithm executed by agent $1$.} \;
\While{$\TRUE$}{
$\mz{1}{k} \gets \nabla \cf_1(\mx{1}{k})~~~~~~~~~~~~~~~~$\hfill \emph{Local computation.}\;
$\ms{k} \gets \left\lceil \textstyle \frac{1}{2}( k + \| \mx{1}{k} \|_2^2 ) \right\rceil ~$ \hfill \emph{Number of consensus steps.} \;
$k_1 \gets k$ \;
$\mave{1}{k} \gets \KwFuncAvg(k_1, \mx{1}{k}, \mz{1}{k}, \ms{k})$  \;
$\mx{1}{k +1} \gets \thresh{\mx{1}{k} - \textstyle \frac{1}{\Lb} \mave{1}{k}}$ \;
 $k \gets k+1$ \;
}
\BlankLine
\BlankLine
\emph{Algorithm executed by agent $p \neq 1$.} \;
\While{$\TRUE$}{
\KwOn({$\funcRecv_p(k_1, \mx{1}{k})$})  { 
   	\If{$k_1 > k$}{
	stop \KwFuncAvg for iter. $k$ \;
	$k \gets k_1$ \;
	$\mx{p}{k} \gets \mx{1}{k_1}$ \;
	$\mz{p}{k} \gets \nabla \cf_p(\mx{p}{k})$ \hfill \emph{Local computation.} \;
	 In next time step, \;
	 ~~~\KwFuncAvg($k_1$, $\mx{1}{k}$, $\mz{p}{k}$)  \;
	}
	}
}
\end{algorithm}

We now detail our CB-DIHT algorithm.
As in DIHT for static networks, each agent has an estimate $\mx{p}{k}$, initially $x_{\textit{init}}$.
For each iteration $k$ of CB-DIHT, agent 1 computes its intermediate vector according to (\ref{eq:z}).
It initiates the diffusive distributed consensus algorithm for iteration $k$ by sending $\mx{1}{k}$ along its outgoing links.  
All agents use the vector $\mx{1}{k}$ as the initiation message for this instance of the consensus algorithm.  
When an agent $p \neq 1$ receives an initiation message containing $\mx{1}{k}$, it updates its local estimate to be identical to 
that of agent 1, i.e., it sets $\mx{p}{k} = \mx{1}{k}$.  It ceases participating in the diffusive consensus algorithm instance for iteration $k-1$ (if applicable).  It then computes $\mz{p}{k}$ according to (\ref{eq:z}), and it begins participating in the diffusive consensus algorithm instance for iteration $k$ of CB-DIHT.

After agent 1 executes $\ms{k} = \lceil (k + \| \mx{1}{k} \|^2)/2 \rceil$ time steps of diffusive distributed consensus, it uses its local estimate of the average, denoted $\mwe{t}$, to compute $\mx{1}{k+1}$ as,
\begin{equation} \label{dynglobal.eq}
\mx{1}{k+1} = \thresh{\mx{1}{k} - \textstyle \frac{1}{\Lb} \mwe{k}},
\end{equation}
where $\Lb > \frac{1}{P} \Lf$.
It then begins a new instance of diffusive distributed consensus for iteration $k+1$ of CB-DIHT. 
Pseudocode for CB-DIHT is given in Algorithm \ref{tvdiht.alg}.

\subsection{Algorithm Analysis} \label{tvan.sec}
\CBDIHT\ requires $O(N)$ storage at each agent.  In each round of diffusive distributed consensus, every agent sends its estimate, an $N$-vector, along all outgoing active links.
 With respect to computational complexity, each agent must compute its local gradient, which, in the case of compressed sensing, consists of matrix-vector multiplication.  Agent 1 performs the thresholding operation which requires a single scan of $\mwe{k}$.

For each iteration $k$, the estimates at agents $p \neq 1$ are identical to those at agent 1.  We note that, due to the time-varying nature of the network, it is possible that some agents may not be initiated in the distributed diffusive consensus instance for a given iteration.  Therefore, the estimates at these agents may skip iterations of IHT.  
By Assumption \ref{network.assum}, each agent's estimate will be updated in infinitely many iterations, and so it suffices to analyze the convergence of the estimate at agent 1.  This estimate evolves as follows,
\begin{eqnarray}
\ms{k} &=&  \left\lceil \textstyle \frac{1}{2}(k + \displaystyle \| \mx{1}{k} \|_2^2) \right\rceil  \label{s.eq} \\
\mwe{k} &=& \sum_{p=1}^P \left[ \Phi(\ms{k})\right]_{1p} \nabla \cf_p(\mx{1}{k}) \label{v.eq} \\
\mx{1}{k+1} &=& \thresh{\mx{1}{k} - \textstyle \frac{1}{\Lb} \mwe{k}}, \label{x2.eq}
 \end{eqnarray}
where $\Phi(\ms{k})$ is the product of the weight matrices for $\ms{k}$ time steps of the diffusive distributed consensus algorithm, i.e., 
\[
\Phi(\ms{k})~=~\Wt{t_{\ms{k}}}~\Wt{t_{\ms{k} - 1}}~\cdots~\Wt{t_{_2}}~\Wt{t_{_1}},
\]
with each $\Wt{t_i}$ satisfying Assumption \ref{weight.assum}.  The notation $[~\cdot~]_{1p}$ indicates the entry of  the matrix 
at row $1$, column $p$.
The vector $\mwe{k}$ is thus agent 1's estimate of the average of the intermediate vectors in iteration $k$ of CB-DIHT.

It is straightforward to show that the evolution of $\mx{1}{k}$ can be formulated as an execution of centralized IHT with inexact gradients.
\begin{proposition} \label{x.prop}
The evolution of the estimate $\mx{1}{k}$ specified by (\ref{s.eq})-(\ref{x2.eq}) can be written as 
\begin{equation}
\mx{1}{k+1} = \thresh{\mx{1}{k} - \textstyle \frac{1}{L}  \left( \nabla f(\mx{1}{k}) + \eps{k} \right)}, \label{newrec.eq}
\end{equation}
where ${L = P \Lb}$, ${\cf(\mx{1}{k}) = \sum_{p=1}^{P} f_p(\mx{1}{k})}$, and  $\eps{k} =  P \mwe{k} - \sum_{p=1}^P \nabla f_p(\mx{1}{k}) $.
\end{proposition}
 \begin{IEEEproof}
By  (\ref{x2.eq}), the vector $\mx{1}{k}$ evolves as, 
 \begin{align*}
 &\mx{1}{k+1} = \thresh{\mx{1}{k} - \textstyle \frac{1}{\Lb} \mwe{k}} \\
 &= \thresh{\mx{1}{k} - \textstyle \frac{1}{\Lb}\left(\frac{1}{P} \nabla f(\mx{1}{k}) + \mwe{k} - \frac{1}{P} \nabla f(\mx{1}{k})\right)} \\
 &= \thresh{\mx{1}{k} - \textstyle \frac{1}{P \Lb} \left( \nabla f(\mx{1}{k}) + P \mwe{k} -  \nabla f(\mx{1}{k})\right)}.
 \end{align*} 
 Substituting with the expressions, $L = P \Lb$ and $\eps{k} =  P \mwe{k} - \sum_{p=1}^P \nabla f_p(\mx{1}{k})$, we obtain
  (\ref{newrec.eq}).
 \end{IEEEproof}

We now show that, under Assumptions \ref{f.assum}, \ref{network.assum}, and \ref{weight.assum},  the sequence of approximation errors is square-summable.
\begin{lemma} \label{eps.lem}
Under Assumptions \ref{f.assum}, \ref{network.assum}, and \ref{weight.assum},
the sequence $\{ \eps{k} \}_{k \geq 0}$ defined in Proposition~\ref{x.prop} satisfies $\sum_{k=0}^{\infty} \| \eps{k} \|_2^2 < \infty$.
\end{lemma}
\begin{IEEEproof}
Let $\mwa{k}$ denote the exact average of the intermediate vectors in iteration $k$ of \CBDIHT, i.e.,
\[
\mwa{t} = \frac{1}{P}\sum_{p=1}^P \nabla \cf_p(\mx{1}{t}).
\]
We can thus express $\eps{k}$ as  
\[
\eps{k} =  P\left(\mwe{k} - \mwa{k}\right).
\]
Using Proposition \ref{avg.thm}, we bound $\| \eps{k} \|_2$ as
\[
\| P( \mwe{k} -  \mwa{k}) \|_2 \leq P \Gamma \gamma^{\ms{k}}  \sum_{p=1}^P \| \nabla \cf_p(\mx{1}{k})\|_2.
\]
Therefore,
\begin{align}
&\sum_{k=0}^{\infty} \|  P(\mwe{k} -  \mwa{k}) \|_2^2  \leq~\sum_{k=0}^{\infty} \left( P \Gamma \gamma^{\ms{k}}  \sum_{p=1}^P \|\nabla \cf_p(\mx{1}{k}) \|_2 \right)^2 \label{sumbound1.eq} \\
&~~~~~~~\leq~P^2 \Gamma^2 \sum_{k=0}^{\infty} \gamma^{2 \ms{k}} ( \gBound \| \mx{1}{k} \|_2 + \hBound)^2, \label{sumbound3.eq}
\end{align}
with $\gBound = \sum_{p=1}^P \gBound_p$ and $\hBound =  \sum_{p=1}^P  \hBound_p$.  Here,  (\ref{sumbound3.eq}) follows from (\ref{sumbound1.eq}) by Assumption \ref{f.assum}(c).

Substituting the value of  $\ms{k}$ from (\ref{s.eq}) into (\ref{sumbound3.eq}), we obtain
\begin{align*}
\sum_{k=0}^{\infty} \| P( \mwe{k} -  \mwa{k}) \|_2^2 &\leq~P^2 \Gamma^2 \gBound^2 \sum_{k=0}^{\infty} \gamma^{k}  \|\mx{1}{k}  \|_2^2 \gamma^{\| \mx{1}{k} \|_2^2} \\
&~~~~+ 2 P^2 \Gamma^2 \gBound \hBound \sum_{k=0}^{\infty} \gamma^k \| \mx{1}{k} \|_2 \gamma^{\| \mx{1}{k} \|_2^2} \\
&~~~~+ P^2 \Gamma^2 \hBound^2 \sum_{k=0}^{\infty} \gamma^{k} \gamma^{\| \mx{1}{k} \|_2^2}.
\end{align*}
We note that since $\gamma \in (0,1)$, the functions $y \gamma^y$, $\sqrt{y} \gamma^y$, and $\gamma^y$ are bounded for any $y \geq 0$ 
(including $y = \| \mx{1}{k} \|^2_2$).
Thus, the sum of the sequence $\{\|  \mwe{k} -  \mwa{k} \|_2^2\}_{k \geq 0}$ is upper-bounded by the sum of a geometric sequence $\{J \gamma^k\}_{k \geq 0}$ for some constant $J$.  Since $\gamma \in (0,1)$, this sequence is summable, and thus $\{\| P( \mwe{k} -  \mwa{k}) \|_2^2\}_{k \geq 0}$ is summable, proving the theorem.
\end{IEEEproof}

The following theorem follows directly from Proposition~\ref{x.prop}, Lemma~\ref{eps.lem}, and Theorems~\ref{lstation.thm} and \ref{lconverge.thm}. 

\begin{theorem} \label{finalconverge.thm}
Let Assumptions \ref{f.assum}, \ref{network.assum}, and \ref{weight.assum} hold, and let $\{\mx{p}{k}\}_{k \geq 0}$, $p=1 \ldots P$, be the sequences generated by {\CBDIHT} with $\Lb>\frac{1}{P} \Lf$.  Then,
\begin{enumerate}
\item Any accumulation point of the sequence $\{\mx{p}{k}\}_{k \geq 0}$, $p=1 \ldots P$, is an $L$-Stationary point of (\ref{opt.eq}).  \item For the compressed sensing problem (\ref{optcs.eq}), with $\text{spark}(A) > K$, the sequences  $\{\mx{p}{k}\}_{k \geq 0}$, $p=1 \ldots P$, converge to an $L$-stationary point.  
\end{enumerate}
Furthermore, if a sequence $\{\mx{p}{k}\}_{k \geq 0}$, converges to an $L$-stationary point $x^*$, then all other sequences, $\{\mx{q}{k}\}_{k \geq 0}$, $q=1 \ldots P, q \neq p$,  converge to $x^*$.
\end{theorem}


\section{Simulations} \label{eval.sec}
In this section, we demonstrate the performance of our distributed algorithms for several recovery problems. 
We also compare our algorithms with previously proposed distributed approaches for sparse signal recovery.
Note that, while \DIHT\  and \CBDIHT\ can be used to recover signals from nonlinear measurements, we are unaware of any other distributed method that addresses this general problem.
Therefore we restrict our evaluations to distributed compressed sensing, for which there are several other existing algorithms.
We now briefly review these other methods.


\subsection{Alternative Algorithms}

As discussed in Section~\ref{intro.sec}, other algorithms for distributed compressed sensing use a convex optimization formulation, for example,  basis pursuit~\cite{CDS98}:
\begin{align}
&\text{minimize}~ \textstyle \frac{1}{P} \sum_{p=1}^P \| x \|_1 \nonumber \\
&\text{subject to}~ \mA{p} x = \mb{p},~~p=1 \ldots P.  \label{convex.eq}
\end{align}
It has been shown that basis pursuit algorithms have recovery guarantees comparable to centralized IHT~\cite{EK12}.
Therefore, distributed basis pursuit and \DIHT\ also have comparable recovery guarantees. 
\CBDIHT\ is based on a generalization of centralized IHT for nonlinear objectives, and the recovery guarantees of this version of IHT 
are not as well studied.  In our evaluations, \DIHT\ and \CBDIHT\ exhibited similar signal recovery capabilities.

In a recent work, Mota et al. compared several distributed basis pursuit algorithms for static networks and showed that \DADMM\ outperformed all other approaches in terms of the number of messages~\cite{MXAP12}.  
We therefore use  \DADMM\ as the representative example in our evaluation, and we repeat the same experiments here. 

For time-varying networks, we compare CB-DIHT with the distributed subgradient algorithm~\cite{LOF11}. This algorithm was proposed to solve a general class of convex optimization problems, of which, (\ref{convex.eq}) is a special case.  Previous work has shown that, in time-varying networks, the subgradient algorithm outperforms the double-looped method in~\cite{JXM11} in similar evaluations.

We briefly describe each of these algorithms.  Pseudocode is Appendix~\ref{code.app}.


\subsubsection{D-ADMM} \DADMM\ is a distributed version of the alternating direction method of multipliers.
The algorithm requires that a graph coloring is available, meaning that every agent is assigned a color such that no two neighboring agents share the same 
color.  
Each agent has its own estimate, $\mx{p}{k}$.  
In every iteration, the agent exchanges an intermediate $N$-vector with all of its neighbors, according to the order dictated by the graph coloring,
and it generates its next iterate by solving a local convex optimization problem. 
In a single iteration of \DADMM, only one color of agents sends messages at a time.  Therefore, one iteration of \DADMM\  takes $c$ times as long as one iteration of distributed consensus. 
We note that, it has not been theoretically verified that D-ADMM converges to the optimal solution of (\ref{convex.eq}) in
general graphs, however its convergence has been demonstrated experimentally~\cite{MXAP12}.  


\subsubsection{Distributed Subgradient Algorithm}

In the distributed subgradient algorithm, each agent $p$ has an estimate $\mx{p}{k}$, initially 0.
Every agent performs a single step of the distributed consensus algorithm to form a weighted average of its and its neighbors'
estimates. Thus, in every iteration, the agent receives an $N$-vector along all of its incoming links in that iteration.
The agent then locally computes $\mx{p}{k+1}$ by performing a projected gradient step with step size $\alpha^{(k)}$.
If the weights used for the consensus algorithm satisfy Assumption \ref{weight.assum} and the step-size sequence $\{ \alpha^{(k)} \}_{k \geq 0}$ is square-summable but not summable,
then, in a static network, this algorithm converges to the optimal solution of (\ref{convex.eq}).  Convergence has also been shown in  time-varying networks that obey Assumption~\ref{network.assum}~\cite{LOF11}.


\begin{table}
\centering
\renewcommand{\arraystretch}{1.3}
\caption{Recovery problem parameters.}  \label{params.tab}
\begin{tabular}{c||ccccc}
\hline
\textbf{Problem}& $N$ & $M$ & $P$ & $K$ & $\lambda_{max}(A \tp A)$ \\
\hline
\textbf{Sparco 902} & 1000 & 200 & 50 & 3 & 1 \\
\textbf{Sparco 7}  & 2560 & 600 & 40  &  20 & 1 \\
\textbf{Sparco 11} & 1024 & 256 & 64  &  32 & $\approx 2283$ \\
\hline
\end{tabular}
\end{table}


\subsection{Evaluation Setup}

We show evaluation results for three compressed sensing problems from the Sparco toolbox~\cite{BFHH07}. 
Details of the problems are given in Table~\ref{params.tab}.
For each problem, we use the measurement matrix $A$ and original, sparse signal $x$ provided by the toolbox.
We generate the measurement vector $b = Ax$ without any measurement noise.  
 For each problem, we divide the measurements (rows from $A$ and $b$) evenly among the agents so that each agent has $M/P$
measurements.

We evaluate each algorithm's performance on five different classes of graphs. 
For each class, we generate five random instances.  The results shown in this section are the averages of the five runs over the five graph instances.  
The first graph type is a Barabasi-Albert (BA) scale free graph~\cite{BA99}.  The second and third are Erd\"{o}s-R\'{e}nyi (ER) random graphs \cite{ER59} where each pair of vertices is connected with probability $pr=0.25$ and probability $pr=0.75$, respectively.
The fourth and fifth graphs are geometric graphs~\cite{P04} with vertices placed uniformly at random in a unit square.  In the fourth graph, two vertices are connected if
they are within a distance  {$d=0.5$} of each other, and in the fifth, vertices are connected if they are within a distance   {$d=0.75$}. 
Of these graphs, the BA graph is the least connected, with 128 edges, on average, for $N=50$ and 171 edges, on average, for $N=64$.
The ER graph with $pr=0.75$ is the most connected, with 992 edges, on average, for $N=50$ and 1,514 edges, on average, for $N=64$.

For simulations in time-varying networks, for each of the graphs described above, we choose ten random subgraphs, ensuring that the union of these subgraphs is the original graph.  We cycle through these ten graphs, one per time step. 

We have implemented all algorithms in Matlab and use CVX~\cite{GB11} to solve the local optimization problems in D-ADMM. 
All algorithms are initiated with each agent's estimate equal to 0.
D-ADMM requires a graph coloring, which we generate using the heuristic from the Matgraph toolbox~\cite{S12}, as in~\cite{MXAP12}. 
While we include the preprocessing phase in our results for \DIHT, we do not include graph coloring preprocessing in our results for \DADMM.
For DIHT, we set $L=2.01$ for Sparco problems 902 and 7.  For Sparco problem 11, we use two values of $L$,
$L = 4570$ and $L=500$.  For CB-DIHT in static networks, we let $\Lb = L/P$, where $L$ is as for DIHT.
For CB-DIHT in time-varying networks, we use $\Lb=2.01/P$ for Sparco problems 902 and 7. For Sparco problem 11, we use
 $\Lb=4570/P$ and $\Lb=600/P$.
For problem 11, the smaller values of $L$ and $\Lb$ are not sufficient to guarantee convergence.  
In our evaluations, for these values,  \DIHT\ and \CBDIHT\ converge to the original signal.
For both \DIHT\ and \CBDIHT, $K$ is set to the value in Table \ref{params.tab}.

For the distributed subgradient algorithm, we experimented with different step-sizes $\alpha^{(k)} = \frac{1}{k^a}$, where $a \in \{0.51, 0.6,  0.7, 0.8, 0.9, 1\}$.
For the most connected graphs, 
the ER graph with $pr=0.75$ and the geometric graph with $d=0.75$, 
the choice of $a$ with the fastest convergence was $0.8$. For the remaining graphs, 
the fastest convergence was with  $a=0.6$.  In the results below, we use $a = 0.7$, which was the value with the second fastest convergence for the vast majority of graphs.
For D-ADMM, we set the algorithm parameters to those that were shown to be best in the same experiments~\cite{MXAP12}.

\begin{table*}
\caption{Signal recovery in a static network for Sparco problem 902 to accuracies of $10^{-2}$ and $10^{-5}$. } \label{sparco902.tab}

\begin{subtable}{1\linewidth}
\centering
\centering
\caption{Total number of values transmitted for each algorithm to converge to within the specified accuracy.} \label{val902.tab}
\renewcommand{\arraystretch}{1.3}
\begin{tabular}{@{}c@{}||cccc||cccc}
 & \multicolumn{4}{c||}{\textbf{Accuracy $10^{-2}$}} & \multicolumn{4}{c}{\textbf{Accuracy $10^{-5} $}} \\
  \cline{2-9}
\begin{tabular}{@{}c} \textbf{Graph} \end{tabular} & 
\renewcommand{\arraystretch}{1} \begin{tabular}{@{}c}~~~\textbf{DIHT}~~ \end{tabular} &  
\renewcommand{\arraystretch}{1} \begin{tabular}{@{}c@{}}~~\textbf{D-ADMM}~~ \end{tabular} &
\renewcommand{\arraystretch}{1} \begin{tabular}{@{}c@{}}~~\textbf{CB-DIHT}~~ \end{tabular} & 
\renewcommand{\arraystretch}{1} \begin{tabular}{@{}c@{}}~~~\textbf{Subgrad.}~~ \end{tabular} &
\renewcommand{\arraystretch}{1} \begin{tabular}{c}~~\textbf{DIHT}~ \end{tabular} &  
\renewcommand{\arraystretch}{1} \begin{tabular}{@{}c@{}}~~\textbf{D-ADMM}~~  \end{tabular} &
\renewcommand{\arraystretch}{1} \begin{tabular}{@{}c@{}}~~\textbf{CB-DIHT}~~  \end{tabular} & 
\begin{tabular}{@{}c@{}}~~~\textbf{Subgrad.}~ \end{tabular}\\
\hline 
\textbf{BA} & \res{2.32}{6}&  
	 \res{2.01}{7}  &  
	 \res{2.27}{8} &$>$\res{5}{10} & 
	\res{5.82}{6} & \res{2.31}{7} & \res{7.78}{8} & $>$\res{5}{10}  \\
\textbf{ER  (\textit{pr}=0.25)}~& \res{2.32}{6}&  
	\res{5.30}{7} &   
	\res{8.45}{8} & \res{2.40}{10} &     
	\res{5.82}{6} & \res{7.31}{7} & \res{3.18}{9} & $>$\res{1}{11}   \\
\textbf{ER (\textit{pr}=0.75)} ~& \res{2.32}{6}&  
	\res{3.26}{8} & 
	\res{1.34}{9} & \res{9.49}{9}  & 
	\res{5.82}{6} & \res{3.85}{8} & \res{6.75}{9} & $>$\res{3}{11}   \\
\textbf{Geo (\textit{d}=0.5)} & \res{2.32}{6}&  
	 \res{3.29}{7} & 
	 \res{1.65}{9} &$\geq$\res{5.3}{9}  &
	\res{5.82}{6}  & \res{3.80}{7} &  \res{3.74}{9} & $>$\res{7}{10}    \\
\textbf{Geo (\textit{d}=0.75)}& \res{2.32}{6}&  
	\res{1.80}{8} & 
	\res{1.38}{9}  & \res{1.11}{10}  &
	 \res{5.82}{6} & \res{2.19}{8} &  \res{5.72}{9} & $>$\res{4}{11}    \\
\hline
\end{tabular}
\vspace{.4cm}
\end{subtable}\\

\begin{subtable}[b]{1\linewidth}
\centering
\caption{Total number of time steps for each algorithm to converge to within the specified accuracy.} \label{time902.tab}
\renewcommand{\arraystretch}{1.3}
\begin{tabular}{@{}c@{}||cccc||cccc}
 & \multicolumn{4}{c||}{\textbf{Accuracy $10^{-2}$}} & \multicolumn{4}{c}{\textbf{Accuracy $10^{-5} $}} \\
  \cline{2-9}
\begin{tabular}{@{}c} \textbf{Graph} \end{tabular} & 
\renewcommand{\arraystretch}{1} \begin{tabular}{@{}c}~~~\textbf{DIHT}~~ \end{tabular} &  
\renewcommand{\arraystretch}{1} \begin{tabular}{@{}c@{}}~~\textbf{D-ADMM}~~ \end{tabular} &
\renewcommand{\arraystretch}{1} \begin{tabular}{@{}c@{}}~~\textbf{CB-DIHT}~~ \end{tabular} & 
\renewcommand{\arraystretch}{1} \begin{tabular}{@{}c@{}}~~~\textbf{Subgrad.}~~ \end{tabular} &
\renewcommand{\arraystretch}{1} \begin{tabular}{c}~~\textbf{DIHT}~ \end{tabular} &  
\renewcommand{\arraystretch}{1} \begin{tabular}{@{}c@{}}~~\textbf{D-ADMM}~~  \end{tabular} &
\renewcommand{\arraystretch}{1} \begin{tabular}{@{}c@{}}~~\textbf{CB-DIHT}~~  \end{tabular} & 
\begin{tabular}{@{}c@{}}~~~\textbf{Subgrad.}~ \end{tabular}\\
\hline 
\textbf{BA} & \res{1.51}{5} & \res{3.14}{5}    & \res{1.02}{5} &$>$\res{4}{8} 
	& \res{3.80}{5} & \res{3.60}{5} & \res{4.39}{6} & $>$\res{4}{8}  \\
\textbf{ER  (\textit{pr}=0.25)}~& \res{1.23}{5} & \res{7.55}{5} & \res{1.58}{6} & \res{7.94}{7} 
	& \res{3.09}{5} & \res{8.63}{5}  & \res{5.61}{6} & $>$\res{4}{8}  \\
\textbf{ER (\textit{pr}=0.75)}~& \res{9.46}{4} & \res{3.14}{6}   & \res{8.36}{5} & \res{2.91}{7} 
	& \res{2.37}{5} & \res{3.71}{6} & \res{3.91}{6}& $>$\res{4}{8}  \\
\textbf{Geo (\textit{d}=0.5)} &  \res{2.22}{5}  & \res{6.92}{5} & \res{4.95}{6} &$\geq$\res{2.8}{7}
	& \res{8.07}{5} &\res{7.98}{5} &  \res{1.09}{7} & $>$\res{4}{8}  \\
\textbf{Geo (\textit{d}=0.75)} & \res{9.46}{4} & \res{2.75}{6} & \res{1.21}{6} & \res{3.40}{7} 
	&  \res{2.37}{5} & \res{3.36}{6} &  \res{4.72}{6} & $>$\res{4}{8} \\
\hline
\end{tabular}
\end{subtable}
\vspace{.3cm}
\end{table*}


\begin{table*} 
\caption{Signal recovery in a static network for Sparco problem 7 to accuracies of $10^{-2}$ and $10^{-5}$. } \label{sparco7.tab}
\begin{subtable}[b]{1\linewidth}
\centering
\caption{Total number of values transmitted for each algorithm to converge to within specified accuracy.}
\renewcommand{\arraystretch}{1.3}
\begin{tabular}{@{}c@{}||cccc||cccc}
 & \multicolumn{4}{c||}{\textbf{Accuracy $10^{-2}$}} & \multicolumn{4}{c}{\textbf{Accuracy $10^{-5} $}} \\
  \cline{2-9}
\begin{tabular}{@{}c} \textbf{Graph} \end{tabular} & 
\renewcommand{\arraystretch}{1} \begin{tabular}{@{}c}~~~\textbf{DIHT}~~ \end{tabular} &  
\renewcommand{\arraystretch}{1} \begin{tabular}{@{}c@{}}~~\textbf{D-ADMM}~~ \end{tabular} &
\renewcommand{\arraystretch}{1} \begin{tabular}{@{}c@{}}~~\textbf{CB-DIHT}~~ \end{tabular} & 
\renewcommand{\arraystretch}{1} \begin{tabular}{@{}c@{}}~~~\textbf{Subgrad.}~~ \end{tabular} &
\renewcommand{\arraystretch}{1} \begin{tabular}{c}~~\textbf{DIHT}~ \end{tabular} &  
\renewcommand{\arraystretch}{1} \begin{tabular}{@{}c@{}}~~\textbf{D-ADMM}~~  \end{tabular} &
\renewcommand{\arraystretch}{1} \begin{tabular}{@{}c@{}}~~\textbf{CB-DIHT}~~  \end{tabular} & 
\begin{tabular}{@{}c@{}}~~~\textbf{Subgrad.}~ \end{tabular}\\
\hline 
\textbf{BA} & \res{6.12}{6} & \res{7.01}{7}  & \res{1.57}{9} &$>$\res{1}{11} 
	& \res{1.64}{7} & \res{1.17}{8} & \res{5.23}{9} & $>$\res{1}{11}  \\
\textbf{ER  (\textit{pr}=0.25)} & \res{6.12}{6} & \res{2.52}{8}  & \res{3.76}{9} &$>$\res{3}{11} 
	& \res{1.64}{7}& \res{3.07}{8} & \res{1.24}{10} & $>$\res{3}{11}   \\
\textbf{ER (\textit{pr}=0.75)}  & \res{6.12}{6} & \res{9.28}{8}  & \res{1.16}{10} & \res{3.63}{10} 
	& \res{1.64}{7} & \res{1.78}{9}  & \res{3.80}{10} & $>$\res{9}{11}   \\
\textbf{Geo (\textit{d}=0.5)} & \res{6.12}{6} & \res{1.23}{8}  & \res{7.45}{9} &$>$\res{1}{11} 
	& \res{1.64}{7} & \res{1.61}{8}  & \res{5.85}{10}& $>$\res{1}{11}    \\
\textbf{Geo (\textit{d}=0.75)} & \res{6.12}{6} & \res{4.53}{8}  &  \res{7.99}{9} & \res{6.25}{10} 
	& \res{1.64}{7} & \res{7.51}{8} & \res{2.65}{10} & $>$\res{1}{12}    \\
\hline
\end{tabular}
\vspace{.4cm}
\end{subtable}\\
\vspace{.4cm}
\begin{subtable}[b]{1\linewidth}
\centering
\caption{Total number of  time steps to within specified accuracy.}
\renewcommand{\arraystretch}{1.3}
\begin{tabular}{@{}c@{}||cccc||cccc}
 & \multicolumn{4}{c||}{\textbf{Accuracy $10^{-2}$}} & \multicolumn{4}{c}{\textbf{Accuracy $10^{-5} $}} \\
  \cline{2-9}
\begin{tabular}{@{}c} \textbf{Graph} \end{tabular} & 
\renewcommand{\arraystretch}{1} \begin{tabular}{@{}c}~~~\textbf{DIHT}~~ \end{tabular} &  
\renewcommand{\arraystretch}{1} \begin{tabular}{@{}c@{}}~~\textbf{D-ADMM}~~ \end{tabular} &
\renewcommand{\arraystretch}{1} \begin{tabular}{@{}c@{}}~~\textbf{CB-DIHT}~~ \end{tabular} & 
\renewcommand{\arraystretch}{1} \begin{tabular}{@{}c@{}}~~~\textbf{Subgrad.}~~ \end{tabular} &
\renewcommand{\arraystretch}{1} \begin{tabular}{c}~~\textbf{DIHT}~ \end{tabular} &  
\renewcommand{\arraystretch}{1} \begin{tabular}{@{}c@{}}~~\textbf{D-ADMM}~~  \end{tabular} &
\renewcommand{\arraystretch}{1} \begin{tabular}{@{}c@{}}~~\textbf{CB-DIHT}~~  \end{tabular} & 
\begin{tabular}{@{}c@{}}~~~\textbf{Subgrad.}~ \end{tabular}\\
\hline 
\textbf{BA} & \res{3.99}{5} & \res{1.32}{6} & \res{6.62}{6} &$>$\res{1}{9} 
	& \res{1.07}{6} & \res{1.72}{6}  & \res{2.13}{7}& $>$\res{1}{9}   \\
\textbf{ER  (\textit{pr}=0.25)} & \res{3.25}{5} & \res{2.98}{6} & \res{6.73}{6} & $>$\res{1}{9} 
	& \res{8.72}{5} & \res{3.63}{6} & \res{2.14}{7} & $>$\res{1}{9}  \\
\textbf{ER (\textit{pr}=0.75)} & \res{2.50}{5} & \res{8.97}{6}  & \res{6.74}{6} & \res{3.93}{7} 
	& \res{6.71}{5} & \res{1.64}{7} & \res{2.15}{7} & $>$\res{1}{9}  \\
\textbf{Geo (\textit{d}=0.5)} & \res{8.49}{5} & \res{2.98}{6} & \res{2.16}{7} & $>$\res{1}{9} 
	& \res{2.28}{6} & \res{3.38}{6}  & \res{4.84}{7} & $>$\res{1}{9}  \\
\textbf{Geo (\textit{d}=0.75)} & \res{2.50}{5} & \res{1.46}{7}  & \res{6.62}{6} & \res{9.82}{7} 
	& \res{6.71}{5} & \res{2.32}{7}  & \res{2.13}{7}& $>$\res{1}{9}  \\
\hline
\end{tabular}
\end{subtable}
\end{table*}


\begin{table*}[t]
\caption{Signal recovery in a static network for Sparco problem 11 to an accuracy of $10^{-2}$. For
DIHT with $L=4750$ and CB-DIHT with $\Lb=4570/P$, in the vast majority of experiments, 
the algorithms converge to an $L$-stationary point that is not the original signal. 
The values shown for DIHT and CB-DIHT are for convergence to the $L$-stationary point;
these values are preceded by a $^\dag$.
For convergence to the original signal, the values in these columns would all be infinite.    
For all other columns, the values shown are for convergence to the original signal.
} \label{sparco11.tab}
\begin{subtable}[b]{1\linewidth}
\centering
\caption{Total number of values transmitted for each algorithm to converge.}
\renewcommand{\arraystretch}{1.3}
\begin{tabular}{@{}c||cccccc}
\begin{tabular}{c} \textbf{Graph} \\ \vspace{.01cm} \end{tabular} & 
\renewcommand{\arraystretch}{1} \begin{tabular}{c}\textbf{DIHT} \\ \scriptsize{$L = 4570$} \end{tabular} &  
\renewcommand{\arraystretch}{1} \begin{tabular}{c}\textbf{DIHT} \\ \scriptsize{$L = 500$} \end{tabular}& 
\renewcommand{\arraystretch}{1} \begin{tabular}{c} \textbf{D-ADMM} \\ \vspace{.01cm} \end{tabular} &
\renewcommand{\arraystretch}{1} \begin{tabular}{@{}c@{}}\textbf{CB-DIHT} \\ \scriptsize{$\Lb = 4570/P$} \end{tabular} & 
\renewcommand{\arraystretch}{1} \begin{tabular}{@{}c@{}}\textbf{CB-DIHT} \\ \scriptsize{$\Lb = 500/P$}\end{tabular}  &  
\begin{tabular}{@{}c@{}} \textbf{Subgradient} \\ \vspace{.01cm} \end{tabular}\\
\hline 
\textbf{BA} & $^\dag$\res{3.37}{7} & \res{1.58}{6} & \res{8.15}{7} & $^\dag$\res{2.27}{10} & \res{3.06}{8}& $>$\res{5}{10}  \\
\textbf{ER (\textit{pr}=0.25)} &  $^\dag$\res{3.37}{7} & \res{1.58}{6}  & \res{5.30}{8} & $^\dag$\res{6.60}{10} & \res{6.96}{8} & $>$\res{1}{11} \\
\textbf{ER (\textit{pr}=0.75)} &  $^\dag$\res{3.37}{7} & \res{1.58}{6} & \res{4.83}{9} &  $^\dag$\res{4.95}{10} & \res{3.66}{9} & \res{1.56}{11} \\ 
\textbf{Geo (\textit{d}=0.5)} &  $^\dag$\res{3.37}{7} &  \res{1.58}{6} & \res{2.26}{8}  &  $^\dag$\res{4.34}{10} & \res{2.54}{10}  & $>$ \res{7}{10} \\
\textbf{Geo (\textit{d}=0.75)} &  $^\dag$\res{3.37}{7} &  \res{1.58}{6}  & \res{2.91}{9}   & $^\dag$\res{7.73}{10} & \res{1.50}{9} & \res{4.02}{10} \\
\hline

\end{tabular}

\vspace{.3cm}
\end{subtable}\\
\vspace{.5cm}
\begin{subtable}[b]{1\linewidth}
\centering
\caption{Total number of time steps for each algorithm to converge.}
\renewcommand{\arraystretch}{1.3}
\begin{tabular}{@{}c||cccccc}
\begin{tabular}{c} \textbf{Graph} \\ \vspace{.01cm} \end{tabular} & 
\renewcommand{\arraystretch}{1} \begin{tabular}{c}\textbf{DIHT} \\ \scriptsize{$L = 4570$} \end{tabular} &  
\renewcommand{\arraystretch}{1} \begin{tabular}{c}\textbf{DIHT} \\ \scriptsize{$L = 500$} \end{tabular}& 
\renewcommand{\arraystretch}{1} \begin{tabular}{c} \textbf{D-ADMM} \\ \vspace{.01cm} \end{tabular} &
\renewcommand{\arraystretch}{1} \begin{tabular}{@{}c@{}}\textbf{CB-DIHT} \\ \scriptsize{$\Lb = 4570/P$} \end{tabular} & 
\renewcommand{\arraystretch}{1} \begin{tabular}{@{}c@{}}\textbf{CB-DIHT} \\ \scriptsize{$\Lb = 500/P$}\end{tabular}  &  
\begin{tabular}{@{}c@{}} \textbf{Subgradient} \\ \vspace{.01cm} \end{tabular}\\
\hline 
\textbf{BA} & $^\dag$\res{1.61}{6} & \res{7.53}{4}  & \res{9.49}{5} & $^\dag$\res{6.76}{7}  & \res{1.27}{6} &$>$\res{4}{8}  \\
\textbf{ER (\textit{pr}=0.25)} & $^\dag$\res{1.39}{6} & \res{6.52}{4} & \res{4.42}{6} & $^\dag$\res{6.75}{7} & \res{1.22}{6}  &$>$\res{4}{8} \\
\textbf{ER (\textit{pr}=0.75)} & $^\dag$\res{1.07}{6} & \res{5.02}{4}   & \res{3.56}{7}  & $^\dag$\res{1.68}{7} & \res{1.16}{6} & \res{1.04}{8} \\ 
\textbf{Geo (\textit{d}=0.5)} & $^\dag$\res{3.11}{6} & \res{1.46}{5} & \res{3.57}{6}  & $^\dag$\res{7.51}{7} & \res{6.96}{7} & $>$\res{4}{8} \\
\textbf{Geo (\textit{d}=0.75)} & $^\dag$\res{1.28}{6}& \res{6.02}{4} & \res{3.29}{7}  & $^\dag$\res{3.66}{7} & \res{1.23}{6}& \res{3.48}{7} \\
\hline
\end{tabular}
\end{subtable}
\vspace{-.4cm}
\end{table*}

\vspace{-.3cm}
\subsection{Results for Static Networks} \label{resultsstatic.sec}

For each algorithm, we measure the number of values sent for $\|\mx{p}{t} - x^*\| / \|x^*\|$ to be less than either $10^{-2}$ or $10^{-5}$ at every agent.
For D-ADMM and the subgradient algorithm, $x^*$ is the original sparse signal from the Sparco toolbox\footnote{We have numerically verified  
that a centralized basis pursuit formulation recovers this original sparse signal.  Additionally, it has been shown in~\cite{MXAP12} that, in the same simulation setup, 
D-ADMM and the distributed
subgradient method converge to the original signal for Sparco problems 902, 7, and 11.}.
\DIHT\ and \CBDIHT\ only guarantee convergence to an $L$-stationary point, and this point may not be the optimal solution to (\ref{optcs.eq}).  We therefore use the relevant $L$-stationary point for $x^*$ where applicable.  
In most experiments,  \DIHT\ and \CBDIHT\ do converge to the original sparse signal. Details of when and how often this occurs are provided below.  For each experiment, we
ran the simulation until convergence within the desired accuracy or for $2 \times 10^5$ iterations, whichever occurred first.

In \DIHT and \CBDIHT, some messages consist of $K$ values and others consists of $N$ values, while in the other algorithms, every message consists of $N$ values.
To standardize the bandwidth comparison between the algorithms, we assume that only one value is sent per message.  Therefore, when an agent sends an $N$-vector
to its neighbor, this  requires $N$ messages.    When an agent sends a $K$-sparse vector, this requires $2K$ messages; 
each component of the vector requires two messages, one containing the index in the vector and one containing the corresponding value.
 Results on the total number of messages that would be sent using a broadcast message model are given in Appendix~\ref{bcast.app}.
  We also measure the time to convergence in a synchronous network where each message is delivered in one time step. For all algorithms, we allow one value to be sent on a given link in each direction per time step. 

We compute the number of of values transmitted by each algorithm as follows.
For \DIHT, each iteration consists of a broadcast phase and convergecast phase.  In the broadcast phase, each agent $p$ sends a $K$-vector to all of its children, requiring $2KQ_p$ messages where $Q_p$ is the number of children that agent $p$ has in the spanning tree (note that $Q_p$ is  less than or equal to the node degree of agent $p$ in the original graph).  In the convergecast phase, each agent, except agent 1, sends an $N$-vector to its parent in the tree, requiring $N$ messages.
In a network of $P$ agents, a spanning tree has $P-1$ edges.  Therefore, for a given problem, \DIHT\ requires the same number of messages for convergence for every network topology.  The number of messages needed to create the spanning tree depends on the network topology, but this messages count is insignificant when compared message count of the algorithm execution.
In  \DADMM\ and the subgradient algorithm, each agent sends an $N$-vector to all of its neighbors in the original graph in each iteration.  Thus, each agent sends 
$\Delta_p N$ messages per iteration, where $\Delta_p$ is the node degree of agent $p$ in the original graph. 
In \CBDIHT, each agent $p$ sends at most $\Delta_p - 1$ activation vectors per iteration, where each activation vector is  $K$-vector that is sent in $2K$ messages.
After activation, an agent sends $N$-vectors to each of its active neighbors in each distributed diffusive consensus round, sending at most $\Delta_p N$ messages
per consensus round.

The results for Sparco problems 902 and 7 are shown in Tables \ref{sparco902.tab} and \ref{sparco7.tab}.   For both problems, \DIHT\ outperforms all other algorithms in both bandwidth and time on all graph instances, and in all cases, \DIHT\ recovers the optimal solution.   
\DIHT\ requires
two orders of magnitude fewer values and time steps than its closest competitor, D-ADMM, to achieve an accuracy of $10^{-2}$. It requires at least one order of magnitude fewer values and time steps than D-ADMM to achieve an accuracy of $10^{-5}$.  
Both CB-DIHT and the subgradient algorithm require more values and time than D-ADMM for these problems.  This indicates that these algorithms pay  a price for tolerating network dynamics even when the network is static.  For an accuracy  of $10^{-5}$, the subgradient algorithm did not converge before the maximum number of iterations.  
The results shown are thus a lower bound on the true number of values and time steps required by this algorithm.  
\CBDIHT\ converged to the optimal solution in all experiments, outperforming the subgradient algorithm by at least one order of magnitude in  bandwidth and time in most cases.

 \begin{table*}    
\caption{Number of iterations of distributed consensus needed for signal recovery in a time-varying network.
For $10^{-5}$ accuracy, the subgradient algorithm required more than $3 \cdot 10^5$ iterations in every instance.
For Sparco problem 11 only,  CB-DIHT with $\Lb=4570/P$ does not always converge to the original signal.  The values shown in this column are for convergence to a sub-optimal $L$-stationary point.  These values are preceded by a $^\dag$.  All other columns give values for convergence to the original signal.} \label{tv.tab}
\centering
\begin{subtable}{.44\textwidth}
\centering
\caption{Sparco problem 902}
\renewcommand{\arraystretch}{1.3}
\begin{tabular}{c||cc||c}
 \multirow{2}{*}{\textbf{Graph}} & \multicolumn{2}{c||}{\textbf{Acc. $10^{-2}$}} &\textbf{Acc. $10^{-5}$} \\
 \cline{2-4} 
& \textbf{CB-DIHT} & \textbf{Subgradient} & \textbf{CB-DIHT}  \\
\hline
\textbf{BA} & \res{2.2}{3}& $>$\res{3}{5} & \res{7.4}{3} \\
\textbf{ER (\textit{pr}=0.25)} & \res{2.5}{3} & $\geq$\res{2.0}{5} & \res{7.3}{3} \\
\textbf{ER (\textit{pr}=0.75)} & \res{1.8}{3} & \res{7.4}{3} & \res{6.0}{3} \\
\textbf{Geo (\textit{d}=0.5)} & \res{6.5}{3} & $>$\res{3}{5} & \res{1.5}{4} \\ 
\textbf{Geo (\textit{d}=0.75)} & \res{1.8}{3} & \res{1.3}{4} & \res{6.0}{3} \\
\hline
\end{tabular}
\end{subtable} 
\hspace{.7cm}
\begin{subtable}{.44\textwidth}
\centering
\caption{Sparco problem 7}
\renewcommand{\arraystretch}{1.3}
\begin{tabular}{c||cc||c}
 \multirow{2}{*}{\textbf{Graph}} & \multicolumn{2}{c||}{\textbf{Acc. $10^{-2}$}} &\textbf{Acc. $10^{-5}$} \\
 \cline{2-4} 
 & \textbf{CB-DIHT} & \textbf{Subgradient} & \textbf{CB-DIHT}  \\
\hline
\textbf{BA} & \res{3.0}{3} & $>$\res{3}{5} & \res{9.3}{3} \\
\textbf{ER} (\textit{pr}=0.25) & \res{2.0}{4} & $>$\res{3}{5} & \res{3.4}{4} \\
\textbf{ER} (\textit{pr}=0.75) & \res{3.3}{3} & \res{4.0}{4} & \res{9.8}{3} \\
\textbf{Geo (\textit{d}=0.5)} & \res{4.9}{4} & $>$\res{3}{5} & \res{7.0}{4} \\ 
\textbf{Geo (\textit{d}=0.75)} & \res{3.5}{3} & \res{6.8}{4}  & \res{1.0}{4} \\
\hline
\end{tabular}
\end{subtable}
\begin{subtable}{1\textwidth}
\vspace{.7cm}
\centering
\caption{Sparco problem 11}
\renewcommand{\arraystretch}{1.3}
\begin{tabular}{c||ccc||cc}
 \multirow{2}{*}{\textbf{Graph}} & \multicolumn{3}{c||}{\textbf{Acc. $10^{-2}$}} &  \multicolumn{2}{c}{\textbf{Acc. $10^{-5}$}} \\
 \cline{2-6}
& \shortstack{\scriptsize{~} \\ \textbf{CB-DIHT} \\ \scriptsize{($\Lb=4750/P$)}} & \shortstack{\textbf{CB-DIHT} \\ \scriptsize{($\Lb=600/P$)}}  &  \shortstack{\textbf{Subgradient} \\ \scriptsize{~}}  & \shortstack{\textbf{CB-DIHT} \\ \scriptsize{($\Lb=4750/P$)}}  & \shortstack{\textbf{CB-DIHT} \\ \scriptsize{($\Lb=600/P$)}}   \\
\hline
\textbf{BA} & $^\dag$\res{6.7}{4} & \res{2.3}{3} &  $>$\res{3}{5} & $^\dag$\res{3.4}{5} & \res{3.9}{3} \\
\textbf{ER (\textit{pr}=0.25)} & $^\dag$\res{7.1}{4} & \res{8.1}{2} &  $>$ \res{3}{5} & $^\dag$\res{1.5}{5} & \res{2.0}{3} \\
\textbf{ER (\textit{pr}=0.75)} & $^\dag$\res{5.8}{4} & \res{7.7}{2} &  $>$\res{3}{5} & $^\dag$\res{1.3}{5} & \res{1.9}{3} \\
\textbf{Geo (\textit{d}=0.5)} & $^\dag$\res{1.0}{5} & \res{5.7}{3} &  $>$\res{3}{5} & $^\dag$\res{3.1}{5} & \res{8.0}{3} \\ 
\textbf{Geo (\textit{d}=0.75)} & $^\dag$\res{5.2}{4}  & \res{8.7}{2} &  $>$\res{3}{5} & $^\dag$\res{1.3}{5} & \res{2.0}{3} \\
\hline
%
\end{tabular}
\end{subtable}

\end{table*}
Results for Sparco problem 11 are shown in Table \ref{sparco11.tab}. 
For DIHT, $L= 4570$ is sufficient to guarantee convergence to an $L$-stationary point. However, convergence  with this $L$ is slower than that of D-ADMM, sometimes requiring up to one order of magnitude more time steps, although with less bandwidth.  
Also, with $L=4570$, DIHT converged to an $L$-stationary point that was suboptimal.  In all simulations, \DIHT\ with $L=500$ converged to the original signal.  In addition, with the smaller $L$, \DIHT\ sent one to three orders of magnitude fewer values than D-ADMM and required one to three orders of magnitude fewer time steps.  The performance of \CBDIHT\ is also significantly worse for $\Lb=4570/P$ than for $\Lb=500/P$, by several orders of magnitude.  Additionally, for $\Lb=4570/P$, \CBDIHT\ converged to a suboptimal $L$-stationary point in all but two graph instances.  With $\Lb=500/P$,  \CBDIHT\ converged to the original signal in all cases.  These results indicate that the bound on $L$ in Theorem~\ref{lstation.thm} is not tight,
 and that further investigation into the convergence conditions for both IHT and its distributed variants is warranted. For the BA graph, the ER graph with $pr=0.25$ and the geometric graph with $d=0.5$, the subgradient algorithm required more than $2 \cdot 10^5$ iterations to converge to within  an error of $10^{-2}$.  
As the table shows, CB-DIHT with the larger $\Lb$ outperformed the subgradient method in both time and bandwidth for all but one graph (Geo $d=0.75$).  
With the smaller value of $\Lb$, CB-DIHT outperformed the subgradient method in both time and bandwidth, usually by at least one order of magnitude.

One interesting observation is that for \DIHT, \CBDIHT, and the distributed subgradient algorithm, as the network connectivity increases, both the bandwidth and convergence time decrease.  In contrast, in \DADMM, as network connectivity increases, the algorithm performance gets worse, requiring both more bandwidth and time.  In the problem formulation used by \DADMM, additional constraints are introduced for each edge in the network graph;  it is our intuition that these additional constraints lead to a decreased convergence rate.  In \DIHT, a more connected the network results in a spanning tree 
with a smaller height; thus, less bandwidth and time is needed to compute each sum.  \CBDIHT\ and the subgradient algorithm both employ distributed consensus algorithms which are well known to converge more quickly in more connected graphs.  We believe that the increase in the convergence rate of the consensus algorithm is carried through to the converge rates  of  \CBDIHT\ and the subgradient algorithm.
  
In \DADMM,  the agents' estimates should become sparse as the algorithm progresses.
If this happens, the agents can exchange sparse representations of their estimates, thus reducing the total number of messages.
We checked for this in our simulations, and for the problems and networks considered herein, using sparse representations had little impact on the evaluation results. 
We believe that it is possible to optimize the communication of D-ADMM in some settings using sparse vector
representations, however, this optimization is beyond the scope of this paper.

Finally, we note that, for some network messaging schemes, such as TDMA, it is more efficient to send messages containing multiple values, rather than a single value per message. 
All algorithms studied in this section would see improvement under this type of scheme.  Since in \DIHT, a $K$-sparse vector is broadcast in each iteration, while the other algorithms always send $N$-vectors, \DIHT\ would not benefit as much.



\subsection{Results for Time-Varying Networks}
We compare \CBDIHT\ with the distributed subgradient method in time-varying networks.  Both algorithms use distributed consensus as a building block;  in the
subgradient method, agents perform one consensus round per iteration, where agents exchange $N$-vectors with their neighbors in that round.  In \CBDIHT, multiple diffusive consensus rounds are performed for each iteration of \DIHT\, and in each consensus round, agents exchange $N$-vectors with their neighbors in that round.
For each algorithm, we count the number of consensus rounds needed for $\|\mx{p}{t} - \xopt \| / \|\xopt \|$ to be less than either $10^{-2}$ or $10^{-5}$ at every agent, where $\xopt$ is as defined in the static network evaluations above.  

The results for time-varying networks are shown in Table \ref{tv.tab}.  We ran each experiment for a maximum of $3 \cdot 10^5$ consensus rounds. 
For the subgradient algorithm, every graph instance required more than $3 \cdot 10^5$ consensus rounds to converge to within $10^{-5}$ of $\xopt$.
Therefore, we do not show these values in the table.  For Sparco problems 902 and 7, \CBDIHT\ converged to the optimal solution in every instance. 
In problem 902, \CBDIHT\ outperformed
the subgradient algorithm by as much as two orders of magnitude for 
an accuracy of $10^{-2}$.  \CBDIHT\ required at least one order of magnitude fewer consensus rounds to achieve an accuracy of $10^{-5}$ in all cases.  For Sparco problem 7, \CBDIHT\
required at least one order of magnitude fewer consensus rounds for both accuracies. 

As before, for \CBDIHT\ on Sparco problem 11, we use a value of $\Lb$ that is sufficient to guarantee convergence, $\Lb=4750/P$, and a smaller value, $\Lb=600/P$, that is not sufficient to guarantee convergence, but nevertheless, converges in all experiments.  For the larger value of $\Lb$, CB-DIHT converged to a suboptimal $L$-stationary point in all but four experiments.  For $\Lb=600/P$, CB-DIHT always converged to the original signal.  For both values of $\Lb$, CB-DIHT required fewer consensus rounds to converge than the subgradient algorithm.  This difference is more pronounced with 
$\Lb=600/P$, where  \CBDIHT\ outperformed the subgradient algorithm by at least two orders of magnitude for both accuracies.   These results reinforce the need for further investigation into the relationship between $\Lb$ and the convergence behavior of  \CBDIHT.


\section{Discussion of Fault Tolerance} \label{discussion.sec}
In both \DIHT\ and \CBDIHT, agent 1 is solely responsible for performing the thresholding operation in each iteration.
A natural question that arises is what happens if this agent fails, or more generally, are these algorithms fault tolerant?

In a discussion of fault tolerance in a static network, we must first assume that the network is synchronous since it is impossible to detect node
failures in an asynchronous network.  Under this assumption, it is straightforward to make \DIHT\ fault tolerant using a self-stabilizing, distributed
algorithm for constructing the spanning tree~\cite{BF03}. The network will autonomously reconstruct the
tree on detection of an agent failure, so long as the underly graph remains connected.  Should the root fail (agent 1), the new root will assume the role of agent 1.  Since the agents all share the same estimate, once the tree is repaired, the algorithm can pick up essentially where it left off.   While the tree is under repair, some agents' estimates may diverge.  However, since there is a single leader and every agent has a single parent in the tree, this hierarchy ensures that the system will return to a consistent state.

After a failure, the objective function will be different; the current estimate serves as warm start for the new optimization problem. 
In addition, if  agent 1 uses $\sum_{p=1} \Lfp$ as its upper bound for $\Lf$ in the original problem, this sum is also an upper bound for $\Lf$
for the problem after the failure.  Therefore, the value of $L$ does not need to change.


In time-varying networks, it is not possible to detect failures.  Therefore, there is no straightforward way to make \CBDIHT\ 
handle the failure of agent 1.  If any other agent fails, \CBDIHT\  can proceed without modification, using the current estimate as the initial estimate for the new optimization problem.   




\section{Conclusion} \label{conclusion.sec}
We have presented two algorithms for in-network, sparse signal recovery based on Iterative Hard Thresholding. 
We first proposed DIHT, a distributed implementation of IHT for static networks that  combines a novel decomposition of centralized IHT with standard tools from distributed computing.
Next, we proposed an extension of DIHT for time-varying networks.  We showed how  centralized IHT can be extended to accommodate inexact computations in each iteration.  We then leveraged these new theoretical results to develop CB-DIHT, a version of DIHT that uses a consensus algorithm to execute these inexact computations in a distributed fashion.  Our evaluations have shown that, in static networks, DIHT outperforms the best-known distributed compressed sensing algorithms in both bandwidth and time by several orders of magnitude.  In time-varying networks, CB-DIHT outperforms the best known algorithm for distributed compressed sensing that accommodates changing network topologies.  We note that, unlike previously proposed algorithms,  both \DIHT\ and \CBDIHT\ can be applied to recovery 
problems beyond distributed compressed sensing, including recovery from nonlinear measurements.  

In future work, we plan to extend our distributed algorithms to support tracking of sparse, time-varying signals.  We also plan to explore the application of \DIHT\ and 
\CBDIHT\  to problems in the Smart Grid.



\appendices

\section{Distributed Computation of $L$ and $\Lb$} 
\label{distL.app}
To determine $L$ or $\Lf$, agent 1 must learn an upper bound on the Lipschitz constant $\Lf$ of the function $\cf(x) = \sum_{p=1}^P \cf_p(x)$.
We first note that, by  Assumption~\ref{f.assum}(b), for all  $x,y \in \mR^N$,
\begin{eqnarray*}
\left\| \sum_{p=1}^P \nabla \cf_p(x) - \sum_{p=1}^P \nabla \cf_p(y) \right\| &\leq& \sum_{p=1}^P  \| \nabla \cf_p(x) - \nabla \cf_p(y) \| \\
&\leq& \sum_{p=1}^P \Lfp \| x - y \|.
\end{eqnarray*}
Therefore, $\sum_{p=1}^P\Lfp$ is an upper bound for $\Lf$.
We now present distributed algorithms by which agent 1 can learn this sum.
\subsubsection*{Computation for \DIHT}
As a pre-processing step for \DIHT, the agents construct a spanning tree of the graph $G$ with agent 1 as its root.
For agent 1 to learn the sum $\sum_{p=1}^{P} \Lfp$, it simply needs to broadcast a request down the tree.
The agents then use a convergecast to aggregate their values for $\Lfp$ up the tree.  The aggregation convergecast is identical to that used to compute the sum of intermediate vectors $\sum_{p=1}^P \mz{p}{k}$ in each iteration of \DIHT, as described in Section~\ref{dihtdesc.sec}.
Once agent 1 knows this sum, it can select an $L > \sum_{p=1}^P \Lfp$ so that the convergence of \DIHT\ is guaranteed.

\subsubsection*{Computation for \CBDIHT}
For \CBDIHT, the step size $\Lb$ must be such that $\Lb > \frac{1}{P}  \Lf$.
One possibility is for the agents to use a distributed consensus algorithm to estimate the average of their respective $\Lfp$
However, agent 1 may not be able to determine how many consensus rounds are  needed to estimate the average with enough accuracy to 
generate a correct upper bound.
A more communication efficient option is for agent 1 to use a distributed algorithm to find $\Lmax = \max\{L_{f_1}, \ldots, L_{f_P}\}$.
Since $\Lmax \geq \frac{1}{P} \sum_{p=1}^P \Lfp$, $\Lmax$ can be used as an (non-strict) upper bound on the average.

To compute  $\Lmax$, every agent stores a variable $m_p$ that it initializes to $\Lfp$.
In every round $t$, the agents sends $m_p$ to all neighbors in that time step.
When an agent receives a value $m_q$ from a neighbor, where $m_q > m_p$, it sets $m_p = m_q$.
If the network satisfies Assumption~\ref{network.assum}, then after at most $2 C \Delta$ rounds of the algorithm,
where $C$ is as defined in Assumption~\ref{network.assum} and $\Delta$ is the diameter of the graph $(V, \Einf)$,
$m_1 = \Lmax$, i.e., agent 1 knows the correct value for $\Lmax$.

 Since the agents do not know $C$ or $\Delta$, agent 1 cannot determine how long to wait before it learns $\Lmax$ and 
 can begin the \CBDIHT\ algorithm.  Instead, the agents execute the distributed max-finding algorithm concurrently with \CBDIHT.
In each iteration $k$ of IHT, agent 1 uses its current value $m_1$ as a bound for the step size, i.e., $\Lb > m_1$.
 After a finite number of time steps, $m_1 = \Lmax$. Therefore, the convergence guarantees stated in 
 Theorems~\ref{lstation.thm}~and~\ref{lconverge.thm} still hold.

\section{Proofs for Iterative Hard Thresholding with Inexact Gradients}
 \label{proofs.app}
In this section, $\| \cdot \|$ denotes the $\ell_2$ norm.

First, for convenience, we restate some relevant results from \cite{B99} and \cite{BE12}.

\begin{lemma}[Descent Lemma] \label{descent.lem}
Let $f$ be a continuously differentiable function whose gradient $\nabla f$ is Lipschitz continuous over $\mR^N$ with constant $\Lf$. 
Then, for every $L \geq \Lf$, 
\[
f(x) \leq h_L(x,y), ~~\text{for all} ~x,y \in \mR^N,
\]
where 
\begin{equation}
h_L(x,y) := f(y) + (x-y)\tp \nabla f(y) + \textstyle \frac{L}{2} \|x - y \|^2. \label{h.eq}
\end{equation}
\end{lemma}

\begin{lemma}[Lemma 2.2 from \cite{BE12}] \label{lstationary.lem}
For any $L > 0$, $\xopt$ is an $L$-stationary point of problem (\ref{opt.eq}) if and only if $\|\xopt\|_0 \leq K$ and,
for $i=1 \ldots N$,
\[
\left| \nabla_i f(\xopt) \right| \left\{ \begin{array}{ll} \leq L\Mk{\xopt} & \text{if}~ \xopt_i = 0 \\
	 = 0 & \text{if}~ \xopt_i \neq 0, \end{array} \right.
\]
where for a given vector $v$, $\Mk{v}$  returns the absolute value of the $K^{th}$ largest magnitude component of $v$.
\end{lemma}

We first derive the following lemma about the relationship between the iterates in $k$ and $k+1$ that are generated by IHT with approximate gradients (analogous to Lemma 2.4 in~\cite{BE12}).
\begin{lemma} \label{fbound.lem}
Let $\xh{0}$ be a $K$-sparse vector, let $\{\xh{k}\}_{k \geq 0}$ be the sequence generated by IHT with inexact gradients in (\ref{ihtin.eq}) with  $L > \Lf$, and let $\cf$ satisfy Assumption \ref{f.assum}. Then, the following inequality holds for all $k \geq 0$:

\begin{align}
f(\xh{k}) - f(\xh{k+1}) &\geq \textstyle \frac{L - \Lf}{2} \| \xh{k} - \xh{k+1}\|^2 \nonumber \\
&~~~~~~~~~ - \left(\xh{k} - \xh{k+1}\right)\tp \eps{k}. \label{flem.eq}
\end{align}
\end{lemma}
\begin{IEEEproof}
Let $C_K$ be the set of $K$-sparse real vectors with $N$ components.  The iteration (\ref{ihtin.eq}) is equivalent to,
\begin{align*}
\xh{k+1}=&\argmin_{v \in C_K} \left\| v -\left(\xh{k} - \textstyle\frac{1}{L} \left( \nabla f(\xh{k}) + \eps{k} \right) \right) \right\|^2 \\
=& \argmin_{v \in C_K} \textstyle \frac{2}{L}(v - \xh{k})\tp \nabla f(\xh{k})  + \textstyle \frac{2}{L} v\tp \eps{k}  \\
&~~~~~~~~~~~ + \|v - \xh{k} \|^2 \\
=& \argmin_{v \in C_K} f(\xh{k}) + (v - \xh{k})\tp \nabla f(\xh{k}) \\
&~~~~~~~~~~~ + \textstyle \frac{L}{2} \|v - \xh{k} \|^2 ~+~ v\tp \eps{k} \\
=& \argmin_{v \in C_K} h_L(v,\xh{k}) ~+~ v\tp \eps{k},
\end{align*}
where $h_L$ is as defined in (\ref{h.eq}).
The above implies that,
\begin{align}
h_L(\xh{k+1}, \xh{k})  + (\xh{k+1} - \xh{k})\tp \eps{k}  \leq&~h_L(\xh{k}, \xh{k})  \nonumber \\
=&~f(\xh{k}). \label{hl.eq}
\end{align}

By Lemma \ref{descent.lem}, we have 
\begin{align}
f(\xh{k}) - f(\xh{k+1})~\geq&~ f(\xh{k}) - h_{\Lf}(\xh{k+1}, \xh{k})   \label{bdl.eq} \\
\geq&~f(\xh{k}) - h_L(\xh{k+1}, \xh{k})  \nonumber  \\
&~~~+ \textstyle \frac{L - \Lf}{2} \| \xh{k+1} - \xh{k} \|^2, \label{dl.eq}
\end{align}
where (\ref{dl.eq}) is obtained from (\ref{bdl.eq}) by applying the identity,
\[
h_{\Lf}(x,y) = h_L(x,y) - \textstyle \frac{L - \Lf}{2} \|x - y\|^2.
\]
Combining (\ref{hl.eq}) and (\ref{dl.eq}), we obtain the result in (\ref{flem.eq}).
\end{IEEEproof}

Using this lemma, we can establish the convergence of the sequence $\{ \| \xh{k} - \xh{k+1}\|^2 \}_{k\geq 0}$, provided 
the sequence of error terms $\{\eps{k}\}_{t \geq 0}$ is square-summable.
\begin{lemma} \label{xconverge.lem}
Let $\xh{0}$ be $K$-sparse, and let $\{\xh{k}\}_{k \geq 0}$ be the sequence generated by IHT with inexact gradients in (\ref{ihtin.eq}) with constant step size $L > \Lf$.  Let the sequence $\{ \eps{k} \}_{k\geq 0}$ be such that $\sum_{k=0}^{\infty} \|\eps{k} \|^2 = E < \infty$.  Then,
\begin{enumerate}
\item There exists a $D < \infty$ such that for all $T \geq 0$,
$\sum_{k=0}^{T}\| \xh{k} - \xh{k+1}\|^2 \leq D$.

\item  $\lim_{k \rightarrow \infty} \| \xh{k} - \xh{k+1}\|^2 = 0$.
\end{enumerate}
\end{lemma}
\begin{IEEEproof}

To prove the first part of the lemma, we show the sequence 
$\{\sum_{k=0}^T  \| \xh{k} - \xh{k+1}\|^2 \}_{T \geq 0}$ is bounded.
Consider the sum over time of $f(\xh{k}) - f(\xh{k+1})$.  We can bound this as,
\begin{align*}
\sum_{k=0}^T \left( f(\xh{k}) - f(\xh{k+1}) \right) &= f(\xh{0}) - f(\xh{T+1})\\
& \leq  f(\xh{0}) - \fBound,
\end{align*}
where the last inequality follows from the fact that $f$ is lower bounded by a constant $\fBound$.
This bound holds for all $T \geq 0$.

Define $A := f(\xh{0}) - \fBound$, and note that since $f(\xh{0})$ is finite, $A$ is also finite. 
By Lemma \ref{fbound.lem}, we have the following for all $T \geq 0$,
\begin{align}
A \geq& \sum_{k=0}^T \left(\textstyle \frac{L - \Lf}{2} \| \xh{k} - \xh{k+1}\|^2 - \left(\xh{k} - \xh{k+1}\right)\tp \eps{k} \right) \label{sum1.eq} \\
\geq& {\textstyle \frac{L - \Lf}{2}}  \sum_{k=0}^T \| \xh{k} - \xh{k+1}\|^2- \sqrt{E \sum_{k=0}^T \|\xh{k} - \xh{k+1}\|^2} \label{sum3.eq},
\end{align}
where (\ref{sum3.eq}) follows from (\ref{sum1.eq}) by the Cauchy-Schwarz inequality and the assumption on the square-summability of the error.

For clarity of notation, let $\beta := \frac{L - \Lf}{2}$, $C := \sqrt{E}$, and $D := \sqrt{\sum_{t=0}^T \|\xh{k} - \xh{k+1}\|^2}$.
We can then rewrite (\ref{sum3.eq}) as,
\begin{equation} \label{quadin.eq}
- \beta D^2   + CD + A  \geq 0.
\end{equation}
Our goal is to show that the sum $D$ is bounded (for all $T \geq 0$), i.e., we must show that every $D \geq 0$ that satisfies (\ref{quadin.eq}) is bounded.
These values are  such that, 
\[
0 \leq D \leq \left(C + \sqrt{C^2 + 4 \beta A}\right)/(2 \beta).
\]
Since $A$ is finite, the sum $D$ is bounded for all $T$, thus proving part one of the lemma.

We now show that ${\{\sum_{t=0}^T\| \xh{k}-\xh{k+1}\|^2 \}_{T \geq 0}}$ converges, implying part two of the lemma (see \cite{R76}, Theorem 3.23). 
To this end, we must show that the sequence is monotonically non-decreasing and bounded.  We have already established that it is bounded. Monotonicity is easily established by,
\begin{align*}
\sum_{k=0}^{T} \| \xh{k} - \xh{k+1}\|^2~=&~ \sum_{k=0}^{T-1} \| \xh{k} - \xh{k+1}\|^2 \\
&~~~~~~~~~~ +  \| \xh{T} - \xh{T+1}\|^2 \\
 \geq&~ \sum_{k=0}^{T-1} \| \xh{k} - \xh{k+1}\|^2.
\end{align*}

\end{IEEEproof}

We now prove the main results of Section \ref{ihtin.sec}.
\begin{restate}{Theorem \ref{lstation.thm}}
\thmLPoint
\end{restate}

\begin{IEEEproof}
Let $\xopt$ be an accumulation point of the sequence of $\{\xh{k}\}_{k \geq 0}$.  
Since the set of $K$-sparse vectors is closed, any such $\xopt$ is $K$-sparse.
If $\xopt$ is an accumulation point, then there exists a subsequence $\{\xh{k_r}\}_{r \geq 0}$ such that $\lim_{r \rightarrow \infty} \xh{k_r} = \xopt$.
By Lemma~\ref{xconverge.lem}, we also have $\lim_{r \rightarrow \infty}\|\xh{k_r}  - \xh{k_r + 1}\|^2 = 0$.
Combing these two statements, we can conclude that $\lim_{r \rightarrow \infty} \xh{k_r + 1} = \xopt$. 

Consider the non-zero components of $\xopt$, i.e., components with $\xopt_i \neq 0$. Since both $\xh{k_r}$ and $\xh{k_r + 1}$ converge to $\xopt$, there exists an $R$ such that  $\xh{k_r}_i, \xh{k_r + 1}_i \neq 0,~\text{for all}~r \geq R$.
Therefore, for $r \geq R$, 
\begin{equation}
\xh{k_r + 1}_i  = \xh{k_r}_i -  \textstyle \frac{1}{L} \left( \nabla_i f\left(\xh{k_r}\right) + \eps{k_r}_i \right). \label{xkr.eq}
\end{equation}
Since $\sum_{k=0}^\infty \| \eps{k} \|^2$ is bounded, we have $\lim_{k \rightarrow \infty} \eps{k}_i = 0$.  Thus, taking $r$ to $\infty$ in (\ref{xkr.eq}), we obtain that $\nabla_i f(\xopt)  = 0$.

Now consider the zero components of $\xopt$, i.e. components with $\xopt_i = 0$.  If there exist an infinite number of indices $k_r$ for which $\xh{k_r+1}_i \neq 0$, then, as before,
\[
\xh{k_r + 1}_i  = \xh{k_r}_i - \textstyle \frac{1}{L}  \left( \nabla_i f\left(\xh{k_r}\right) + \eps{k_r}_i \right), 
\]
which implies $\nabla_i f(\xopt)  = 0$, and thus $| \nabla_i f(\xopt) | \leq L \Mk{\xopt}$.  If there exists a $Q > 0$ such that for all $r > Q$, $\xh{k_r + 1}_i = 0$, then,
\begin{align*}
&\left| \xh{k_r}_i - \textstyle \frac{1}{L}\left(\nabla_i f(\xh{k_r}) + \eps{k_r}_i \right) \right| \\
&~~~~~~~~~~~~~\leq~ \Mk{\xh{k_r} - \textstyle \frac{1}{L} \left( \nabla f\left(\xh{k_r}\right) + \eps{k_r}\right)}  \\
&~~~~~~~~~~~~~=~ \Mk{\xh{k_r + 1}}.
\end{align*}
Taking $r$ to infinity and noting that $\eps{k_r} \rightarrow 0$, we obtain,
\[
\left| \xopt_i - \textstyle \frac{1}{L} \nabla_i f(\xopt) \right| \leq \Mk{\xopt},
\]
or equivalently, $\left| \nabla_i f(\xopt) \right| \leq L \Mk{\xopt}$.  Therefore, by Lemma \ref{lstationary.lem}, $\xopt$ is an $L$-stationary point,  proving the theorem.
\end{IEEEproof}

\begin{restate}{Theorem \ref{lconverge.thm}}
\thmCS
\end{restate}
\begin{IEEEproof}
First, we show that the sequence $\{\xh{k}\}_{k\geq 0}$ is bounded.  
Applying Lemma \ref{fbound.lem}, we have
\begin{align}
&\sum_{s=0}^{k-1} f(\xh{s}) - f(\xh{s+1}) \geq \sum_{s=0}^{k-1}\textstyle \frac{L - \Lf}{2} \| \xh{s} - \xh{s+1}\|^2 \nonumber \\
&~~~~~~-~  \sum_{s=0}^{k-1} \|\xh{s} - \xh{s+1}\| \|\eps{s}\|  \\
&f\left(\xh{0}\right) - f\left(\xh{k}\right) \geq  \sum_{s=0}^{k-1} \textstyle \frac{L - \Lf}{2} \| \xh{s} - \xh{s+1}\|^2 \nonumber \\
&~~~~~~-~ \sqrt{\sum_{s=0}^{k-1} \| \eps{s} \|^2} \sqrt{\sum_{s=0}^{k-1} \|\xh{s} - \xh{s+1}\|^2} \label{fdiff.eq}
\end{align}
By the assumption on the square summability of the error terms, there exists $E < \infty$ such that $E = \sum_{s=0}^{k-1} \| \eps{s} \|^2$. 
We also define $0 \leq F < \infty$ such that ${F = \sum_{s=0}^{k-1}\|\xh{s}-\xh{s+1}\|^2}$.
Note that, by Lemma \ref{xconverge.lem}, such an $F$ exists.
With these definitions,  we arrive at the following inequality for $k \geq 0$,
\[
f(\xh{0}) + \sqrt{E} \sqrt{F} \geq f(\xh{k}).
\]
Let  $T$ be the level set,
\[
T = \left\{ x \in \mR^N : f(x) \leq f(\xh{0}) + \sqrt{EF} \right\},
\]
and note that the sequence $\{\xh{k}\}_{k\geq 0}$ is contained in this set.
For $f(x) = \|A x - b\|_2^2$, if $x$ is $K$-sparse and $\text{spark}(A) > K$, then the set $T$ is bounded (see~\cite{BE12}, Theorem 3.2).  
Thus the sequence $\{\xh{k}\}_{k\geq 0}$ is bounded.

Using an argument identical to one in the proof of Theorem 3.2 in~\cite{BE12}, it can be shown that $\{\xh{k}\}_{k\geq 0}$ converges to
an $L$-stationary point. 
We repeat this argument here for completeness.
The boundedness of $\{\xh{k}\}_{k\geq 0}$  implies that there exists a subsequence $\{\xh{k_j}\}_{j \geq 0}$ that converges to an $L$-stationary point $x^*$.  
By Lemma 2.1 in~\cite{BE12}, there are only a finite number of $L$-stationary points.  
Assume that the sequence $\{\xh{k}\}_{k \geq 0}$ does not converge to $x^*$.
This means that there exists an $\epsilon_1 > 0$ such that for all $J \geq 0$, there exists a $j > J$ with $\|\xh{j} - x^* \| \geq \epsilon_1$.

Define $\epsilon_2 > 0 $ to be less than the minimum distance between all pairs of $L$-stationary points,
and define $\epsilon = \min\left(\epsilon_1, \epsilon_2\right)$. 
Without loss of generality, assume that $\{\xh{k_j}\}_{j \geq 0}$ satisfies $\| \xh{k_j} - x^*\| \leq \epsilon$ for all $j \geq 0$,
and define the sequence $\{\xh{t_j}\}_{j \geq 0}$ with,
\[
t_j = \max\{l: \|x^{(i)} - x^*\| \leq \epsilon, i=k_j, k_j + 1, \ldots , l\}.
\]
Since, by assumption, $\{\xh{k}\}_{k \geq 0}$ does not converge to $x^*$, each $t_j$ is well-defined. 
Given the definition of $\xh{t_j}$ and the fact that $\| \xh{k} - \xh{k+1} \| \rightarrow 0$ (by Lemma \ref{xconverge.lem}), there exists a subsequence of $\xh{t_j}$ 
that converges to a point $z^*$ with ${\|x^*-z^*\|\leq\epsilon}$.
The existence of this subsequence contradicts the assumption that $\epsilon$ was chosen so every accumulation 
point $y \neq x^*$ is such that $\|x^* - y \| >\epsilon$. Therefore, the sequence $\{\xh{k}\}_{k \geq 0}$ converges to $x^*$.

\end{IEEEproof}

\section{Pseudocode for Diffusive Distributed Consensus}
\label{diffcon.app}

\begin{algorithm}[h]
\caption{Diffusive Distributed Consensus algorithm for agent $p$.}
\label{diffcon.alg}
\SetAlgoNoLine
\SetAlgoNoEnd
\DontPrintSemicolon
\SetNoFillComment
\KwInit{
\If{$p=1$}{
$\textit{initiated} \gets \TRUE$ \;
}
\Else{
$\textit{initiated} \gets \FALSE$ \;
}
$\textit{activeNeighbors} \gets \emptyset$ \;
$\mv{p}{0} \gets $ initial value \; 
}
\BlankLine 
\For{$t=0 \ldots \infty$}{
	\If{$\textit{initiated} = \TRUE$}{
		$\mathcal{N}_p(t) \gets$ all agents $q \in \textit{activeNeighbors}$ \;
			~~~~~~where $(p,q) \in \Et{t}$ \;
		$\mv{p}{t+1} \gets  \sum_{q \in \mathcal{N}_p(t) }^P \wt{p}{q}{t}~\mv{q}{t}$ \;

		\For{$(q,p) \in \Et{t}$ and $q \notin \textit{activeNeighbors}$}{
			send \INITIATE\ to $q$ \;
		 	$\textit{activeNeighbors} \gets \textit{activeNeighbors} \cup \{q\}$ \;
		}
	} 
	\If{receive \INITIATE\ from some agent $q$}{
		\If{$\textit{initiated} = \FALSE$}{
			$\textit{initiated} \gets \TRUE$ \;
		}
		$\textit{activeNeighbors} \gets \textit{activeNeighbors} \cup \{q\}$ \;
	}
}
\end{algorithm}

Pseudocode for the diffusive distributed consensus algorithm is given Algorithm \ref{diffcon.alg}.


\begin{algorithm}[t]
\caption{D-ADMM for agent $p$ with color $c$.  Here $D_p$ denotes the node degree of agent $p$.}
\label{dadmm.alg}
\SetAlgoNoLine
\SetAlgoNoEnd
\DontPrintSemicolon
\SetNoFillComment
\KwInit{
$\mx{p}{0} \gets 0$ \;
$\mg{p}{0} \gets 0$ \;
$k \gets 0$ \;
}
\BlankLine
\While{$\TRUE$}{
	\KwOn(receive $\mx{q}{k+1}$ from neighbors with lower colors){
	$u_p^{(k)} \gets  \mg{p}{k} - \displaystyle \rho\sum_{\substack{q \in \cN{p} \\ col(q) < c}} \mx{q}{k+1}-\rho\sum_{\substack{q \in \cN{p} \\ col(q) > c}} \mx{q}{k}  $ \;
	$\mx{p}{k+1} \gets \argmin_{x_p} \frac{1}{P} \| x_p \|_1 + {u_p^{(k)}}\tp  x_p + \frac{D_p}{2}\|x_p\|_2^2$  \;
	$~~~~~~~~~~~~~\mbox{subject to}~ \mA{p}x_p = \mb{p}$ \;
	send $\mx{p}{k+1}$ to all neighbors \;
	}
	\KwOn(receive $\mx{q}{k+1}$ from all neighbors){
		$\mg{p}{k+1} \gets \mg{p}{k} + \rho \sum_{q \in \cN{p}} \left(\mx{p}{x+1} - \mx{q}{x+1}\right)$ \;
	}
	$k \gets k+1$ \;
}
\end{algorithm}

\begin{algorithm}
\caption{Distributed subgradient algorithm.}
\label{subgrad.alg}
\SetAlgoNoLine
\SetAlgoNoEnd
\DontPrintSemicolon
\SetNoFillComment
\KwInit{
$\mx{p}{0} \gets 0$ \;
$k \gets 0$ \;
}
\BlankLine
\While{$\TRUE$}{
  $u_p^{(k)} \gets \sum_{q=1}^P [W(k)]_{pq} \mx{q}{k}$ \;
  $g_p^{(k)} \gets~\mbox{subgradient of}~\|x_p\|_1~\mbox{at}~\mx{p}{k}$ \;
  $\mx{p}{k+1} \gets \left[ u_p^{(k)}  - \alpha^{(k)} g_p^{(k)} \right]^{+}_{\mA{p} x_p = \mb{p}}$ \;
}
\end{algorithm}

\section{Pseudocode for Other Recovery Algorithms} \label{code.app}
The pseudocode for D-ADMM is given in Algorithm \ref{dadmm.alg}.  The pseudocode for the distributed subgradient algorithm is given in Algorithm \ref{subgrad.alg}.
In the distributed subgradient algorithm, $[~\cdot~]^{+}_{\mA{p} x_p - \mb{p}}$ denotes the projection  onto the constraint set $\mA{p} x_p = \mb{p}$.

\begin{table*}
\caption{Total number of broadcasts needed for convergence to an accuracy of $10^{-2}$ in a static network.} \label{broadcast.tab}
\begin{subtable}{.49\linewidth} \label{broadcast902}
\centering
\caption{Sparco problem 902.}
\renewcommand{\arraystretch}{1.3}
\begin{tabular}{@{}c||cccc}
 \textbf{Graph} & 
\renewcommand{\arraystretch}{1} \textbf{DIHT} &  
\renewcommand{\arraystretch}{1} \textbf{D-ADMM} &
\renewcommand{\arraystretch}{1} \textbf{CB-DIHT}  & 
\renewcommand{\arraystretch}{1} \textbf{Subgrad.} \\
\hline 
\textbf{BA} &  \res{2.31}{6} 
& \res{3.97}{6} &  \res{5.10}{7} &  $>$\res{1.00}{10}  \\
\textbf{ER (\textit{pr}=0.25)} & \res{2.31}{6}  
 & \res{5.24}{6} & \res{7.89}{7} & \res{1.99}{9} \\
\textbf{ER (\textit{pr}=0.75)} & \res{2.30}{6}  
& \res{8.70}{6} & \res{4.18}{7} & \res{9.54}{8} \\ 
\textbf{Geo (\textit{d}=0.5)} &  \res{2.31}{6} 
& \res{4.43}{6} & \res{2.48}{8} & \res{1.91}{9} \\
\textbf{Geo (\textit{d}=0.75)} & \res{2.30}{6} 
& \res{6.85}{6} & \res{6.05}{7} & \res{8.50}{8} \\
\hline
\end{tabular}
\vspace{.2in}
\end{subtable} 
\begin{subtable}{.49\linewidth} \label{broadcast7.tab}
\centering
\caption{Sparco problem 7.} 
\renewcommand{\arraystretch}{1.3}
\begin{tabular}{@{}c||cccc}
 \textbf{Graph} & 
\renewcommand{\arraystretch}{1} \textbf{DIHT} &  
\renewcommand{\arraystretch}{1} \textbf{D-ADMM} &
\renewcommand{\arraystretch}{1} \textbf{CB-DIHT}  & 
\renewcommand{\arraystretch}{1} \textbf{Subgrad.} \\
\hline 
\textbf{BA} &  \res{6.03}{6} 
& \res{1.71}{7} & \res{3.31}{8} & $>$\res{2.56}{10}   \\
\textbf{ER (\textit{pr}=0.25)} &  \res{6.02}{6} 
& \res{2.07}{7} & \res{3.36}{8} & $>$\res{2.56}{10}  \\
\textbf{ER (\textit{pr}=0.75)} & \res{6.02}{6} 
& \res{2.52}{7} & \res{3.37}{8} & \res{9.83}{8} \\ 
\textbf{Geo (\textit{d}=0.5)} &  \res{6.03}{6} 
 & \res{1.65}{7} & \res{1.08}{9} & $>$\res{2.56}{10} \\
\textbf{Geo (\textit{d}=0.75)} &  \res{6.02}{6}  
& \res{3.40}{7} & \res{3.31}{8} & \res{2.45}{9} \\
\hline
\end{tabular}
\vspace{.2in}
\end{subtable} 
\begin{subtable}[b]{1\linewidth} \label{broadcast11.tab}
\centering
\caption{Sparco problem 11. For
DIHT with $L=4750$ and CB-DIHT with $\Lb=4570/P$, in the vast majority of experiments, 
the algorithms converge to an $L$-stationary point that is not the original signal. 
The values shown for DIHT and CB-DIHT are for convergence to the $L$-stationary point;
these values are preceded by a $^\dag$.
For convergence to the original signal, the values in these columns would all be infinite.    
For all other columns, the values shown are for convergence to the original signal.
}
\renewcommand{\arraystretch}{1.3}
\begin{tabular}{@{}c||cccccc}
\begin{tabular}{c} \textbf{Graph} \\ \vspace{.01cm} \end{tabular} & 
\renewcommand{\arraystretch}{1} \begin{tabular}{c}\textbf{DIHT} \\ \scriptsize{$L = 4570$} \end{tabular} &  
\renewcommand{\arraystretch}{1} \begin{tabular}{c}\textbf{DIHT} \\ \scriptsize{$L = 500$} \end{tabular}& 
\renewcommand{\arraystretch}{1} \begin{tabular}{c} \textbf{D-ADMM} \\ \vspace{.01cm} \end{tabular} &
\renewcommand{\arraystretch}{1} \begin{tabular}{@{}c@{}}\textbf{CB-DIHT} \\ \scriptsize{$\Lb = 4570/P$} \end{tabular} & 
\renewcommand{\arraystretch}{1} \begin{tabular}{@{}c@{}}\textbf{CB-DIHT} \\ \scriptsize{$\Lb = 500/P$}\end{tabular}  &  
\begin{tabular}{@{}c@{}} \textbf{Subgradient} \\ \vspace{.01cm} \end{tabular}\\
\hline 
\textbf{BA} &  $^\dag$\res{3.18}{7}  &  \res{1.49}{6} 
& \res{1.52}{7} & $^\dag$\res{4.33}{9} & \res{8.14}{7} & $>$\res{1.31}{10} \\
\textbf{ER (\textit{pr}=0.25)} &   $^\dag$\res{3.18}{7} & \res{1.49}{6} 
 & \res{3.45}{7} & $^\dag$\res{4.32}{9} & \res{7.84}{7} & $>$\res{1.31}{10} \\
\textbf{ER (\textit{pr}=0.75)} &  $^\dag$\res{3.18}{7}  & \res{1.48}{6} 
& \res{1.04}{8} & $^\dag$\res{1.08}{9} & \res{7.41}{7} & \res{3.34}{9} \\ 
\textbf{Geo (\textit{d}=0.5)} &   $^\dag$\res{3.18}{7} & \res{1.49}{6} 
& \res{2.28}{7} & $^\dag$\res{6.31}{8} & \res{4.46}{9} & $>$\res{1.31}{10} \\
\textbf{Geo (\textit{d}=0.75)} &  $^\dag$\res{3.18}{7} & \res{1.48}{6} 
& \res{8.04}{7} & $^\dag$\res{2.34}{9} & \res{7.85}{7} & \res{1.11}{9} \\
\hline
\end{tabular}
\end{subtable}
\end{table*}
\section{Simulation Results for Broadcast Communication} \label{bcast.app}

We present evaluation results for the convergence of the different algorithms using a broadcast model of communication in a static network.
We use the same evaluation setting as in Section~\ref{resultsstatic.sec}.  We note that the time evaluation results in Section~\ref{resultsstatic.sec} also apply to
a broadcast setting since they account for messages being sent on an agent's links in parallel.

When an agent broadcasts a message, it is sent to all of its neighbors in the original graph.  In the broadcast phase of a 
\DIHT\ iteration, when an agent broadcasts a message, all of its neighbors receive the message, but only the agent's children process this message;
the others discard it.  Similarly, in the convergecast phase, while all neighbors of an agent receive a broadcast message, only the parent of
that agent processes it.  
In \CBDIHT, \DADMM, and the subgradient method, each message is broadcast to all of the agents' neighbors in the original graph, and they all
process that message. 
In all algorithms, each broadcast contains a single value.  So, to send an $N$-vector, $N$ broadcasts are needed, and to send
a $K$-vector, $2K$ broadcasts are needed.  

For each algorithm, we measure the number of broadcasts needed for $\|\mx{p}{t} - x^*\| / \|x^*\|$ to be less than $10^{-2}$ at all agents.
As in Section~\ref{eval.sec}, for D-ADMM and the subgradient algorithm, $x^*$ is the original sparse signal from the Sparco toolbox.
For \DIHT\ and \CBDIHT, $x^*$ is the relevant $L$-stationary point.  We clearly indicate when $x^*$ is not the original signal in the evaluation results. 
For each experiment, we
ran the simulation until convergence within the desired accuracy or for $2 \times 10^5$ iterations, whichever occurred first.  

The simulation results are given in Table~\ref{broadcast.tab}.  In \DIHT\, the total number of broadcasts per iteration depends on the topology of the spanning tree,
since leaf nodes do not have any children to which to broadcast the iterate.  The topology of the tree, in turn, depends on the topology of the original graph.
 Therefore, while the total number of broadcasts is similar for each graph for a given problem, it is not identical.
  For Sparco problems 902 and 7, \DIHT\ requires significantly fewer broadcasts than \DADMM,
and this difference is more pronounced in problem 7.  \CBDIHT\ requires an order of magnitude more broadcasts than \DIHT\ and \DADMM\ in most cases.
 and the subgradient algorithm requires at least two orders of magnitude more broadcasts than \DIHT\ and \DADMM\ in all cases.
 For Sparco problem 11, when $L = 4570$, \DIHT\ requires more broadcasts than \DADMM\ to converge, and it does not converge to the original signal.
 When $L=500$, \DIHT\ recovers the original signal, requiring an order of magnitude fewer broadcasts than \DADMM\ to do so.  Both \CBDIHT\ and the 
 subgradient algorithm require significantly more broadcasts to achieve convergence.


\bibliographystyle{IEEEtran}
\bibliography{cs}






\end{document}